\documentclass[aip,jcp,reprint,groupedaddress,longbibliography]{revtex4-2}

\usepackage{amsmath} 
\usepackage{graphicx} 
\usepackage{float}
\usepackage[version=4]{mhchem}

\renewcommand{\vec}{\bf}

\begin{document}

\title{How to validate machine-learned interatomic potentials}

\author{Joe D. Morrow}
\author{John L. A. Gardner}
\author{Volker L. Deringer}
\email{volker.deringer@chem.ox.ac.uk}
\affiliation{Department of Chemistry, Inorganic Chemistry Laboratory, University of Oxford, Oxford OX1 3QR, UK}

\begin{abstract}
    Machine learning (ML) approaches enable large-scale atomistic simulations with near-quantum-mechanical accuracy. 
    With the growing availability of these methods there arises a need for careful validation, particularly for physically agnostic models -- that is, for potentials which extract the nature of atomic interactions from reference data.
    Here, we review the basic principles behind ML potentials and their validation for atomic-scale materials modeling.
    We discuss best practice in defining error metrics based on numerical performance as well as physically guided validation.
    We give specific recommendations that we hope will be useful for the wider community, including those researchers who intend to use ML potentials for materials ``off the shelf''.
\end{abstract}

\maketitle

\section{Introduction}

Machine learning (ML) based interatomic potentials are becoming increasingly popular for mainstream materials modeling. \cite{Behler2017, Deringer2019, Noe2020, Unke2021, Friederich2021}
ML potentials are ``trained'' using quantum-mechanical reference data (energies and forces on atoms) and, once developed and properly validated, enable large-scale atomistic simulations at a similar level of quality whilst requiring only a small fraction of the computational cost.
In recent years, ML potentials have been used to address fundamental research questions that would otherwise have been inaccessible for quantum-accurate studies: the complex high-pressure phase behavior of seemingly simple elements,\cite{Cheng2020, Deringer2021, Zong2021} the microscopic growth mechanism of amorphous carbon films, \cite{Caro2018} 
or the photodynamics of a biologically relevant molecule. \cite{Westermayr2022} 
At the same time, ML potentials are being developed for diverse functional materials, with applications including phase-change chalcogenides, \cite{Sosso2013, Konstantinou2019} battery electrodes and solid-state electrolytes, \cite{Artrith2018, Wang2020b, Wang2022, Staacke2022} and multicomponent alloys.\cite{Gubaev2019, Marchand2022} The field is thriving, without any doubt.

With ML potentials becoming increasingly available, it becomes important to ensure careful validation of both the overall methods and the specific models.
This requirement is particularly important for machine-learned, data-driven models, which do not have (much) physical information ``built in'' by construction.
A key step in validation was reported by Zuo et al., who performed a detailed study of numerical energy and force prediction errors across different types of ML potentials. \cite{Zuo2020} The focus of that work was on the fair comparison on different fitting frameworks, and on the identification of a ``Pareto front'' of efficiency, i.e., of the respective most accurate methods and settings at a given level of computational cost. 
The validation tests used by Zuo et al.\ included liquid and crystalline structures for a number of elements, \cite{Zuo2020} and an advantage of this protocol is that it could, in principle, be applied to any elemental system.
In contrast, a recent publication on a carbon ML potential includes various more subject-specific (or ``domain-specific'') tests, such as the formation energies of specific topological defects in graphene, \cite{Rowe2020} and extended benchmarks and more complex tests were later carried out by other groups. \cite{Qian2021, Aghajamali2021, Qamar2022}
Earlier already, de Tomas et al.\ had stressed the importance of careful property-based validation for carbon potentials, including for an ML-based example as well as established empirical potentials\cite{DeTomas2016, DeTomas2019} -- indeed, many questions around validation are relevant for {\em any} type of interatomic potential, machine-learned or otherwise.
Finally, the importance of validating ML potentials based not only on numerical values, but also on the predicted physical behavior, is being pointed out increasingly in the literature. \cite{George2020, Kovacs2021, Fu2022}

The aim of the present work is to review and discuss validation criteria for ML potentials in a ``tutorial'' style. It covers both numerical 
and physically-guided validation, whilst being clearly focused on ML potential models for atomistic simulations. We aim to complement related works by others: more generally on validation for ML models of various types, \cite{Vishvakarma2021} and recently published general best-practice guidelines for the use of ML in chemistry. \cite{Artrith2021, Bender2022}
Our work focuses strongly on inorganic {\em materials}, and we refer the reader to an excellent tutorial on neural-network potentials which places stronger emphasis on the construction of {\em molecular} force fields -- for example, for water. \cite{Miksch2021}
In what follows, we provide a brief overview of relevant methods and concepts, illustrative examples, and best-practice recommendations.

\begin{figure*}
    \centering
    \includegraphics[width=17.8cm]{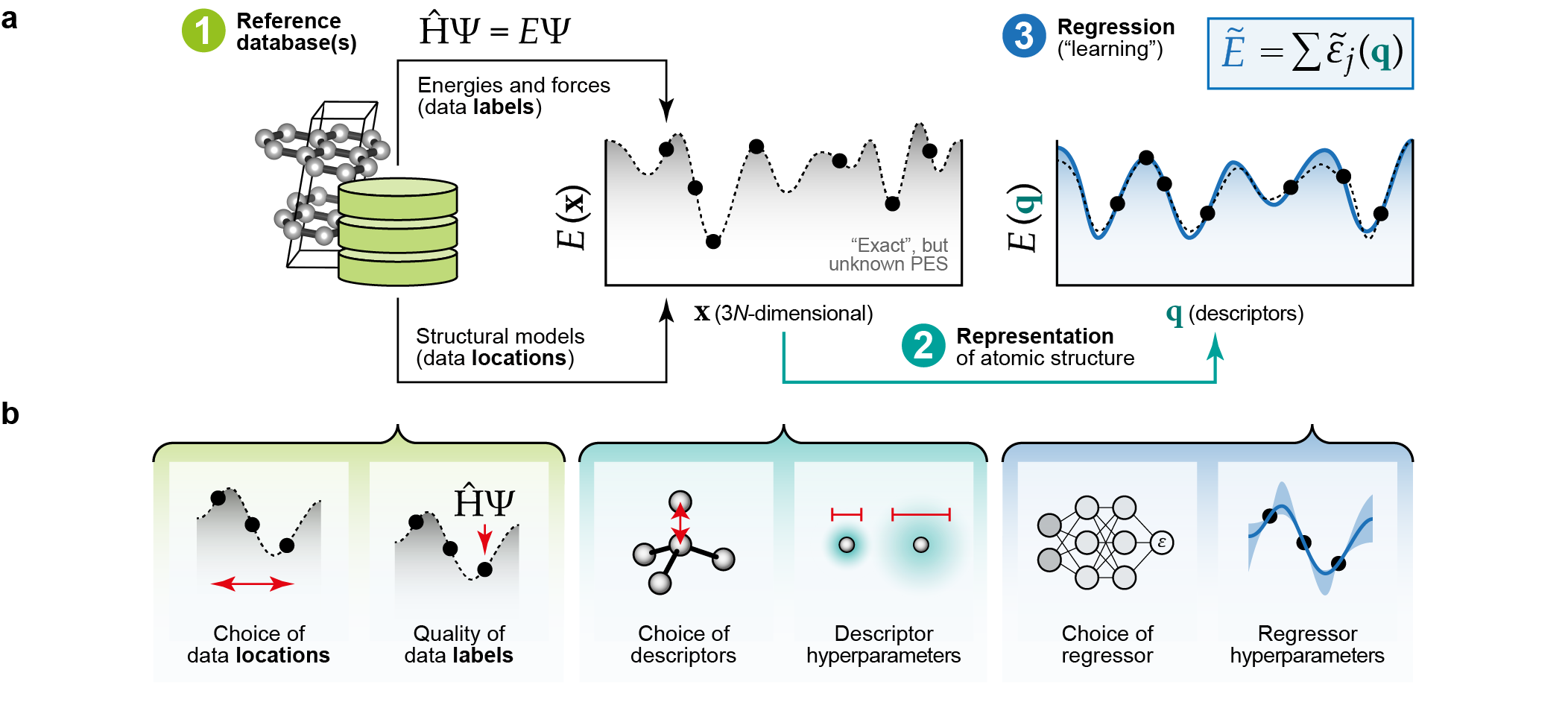}
    \caption{Machine learning interatomic potentials.
    (\textbf{a}) General overview of the methodology, with the three main ``ingredients'' highlighted.\cite{Deringer2019, Deringer2020b} 
    The database (point 1) contains representative structural models which are labeled with ground-truth energies and forces (typically obtained using DFT).
    The representation encodes the atomic environment, expressed by the general vector {\vec{q}}. 
    The regression task creates the model, $\tilde{E}$, and in current ML potential models it is common to predict the total energy as a sum of machine-learned local energies, $\tilde{\varepsilon}_{j}$.
    (\textbf{b}) Criteria and choices that affect the quality of an ML potential, indicated by highly schematic sketches. The color-coding corresponds to panel (a) and indicates effects of reference databases in green, representations (descriptors) in turquoise, and the regression framework in blue, respectively. 
    Panel (a) is adapted from earlier overview articles in Ref.\ \citenum{Deringer2019} (Copyright 2019 WILEY‐VCH Verlag GmbH \& Co.\ KGaA, Weinheim) and Ref.\ \citenum{Deringer2020b} (published under a Creative Commons Attribution license, http://creativecommons.org/licenses/by/4.0/).
    }
    \label{fig:MLP_overview}
\end{figure*}

\section{What are ML potentials?}

To set the scene, we start by briefly reviewing what ML potentials are in the first place. We will focus on the major types of choices to be made in developing ML potentials, and how these choices affect the quality of the model. A reader who is familiar with the methodology may wish to skip to the next section.

Machine learning means extracting information from large datasets -- in this case, from quantum-mechanical energies and forces. An ML potential is therefore a complex mathematical model for a given potential-energy surface (PES) whose parameters have been learned from reference data (the {\em ground truth}). There are three main ingredients for doing so, \cite{Deringer2019} as illustrated in Fig.\ \ref{fig:MLP_overview}a. 

The first step in constructing an ML potential is building the reference database to which the fit is made:
devising and selecting small-scale structural models that contain enough ``relevant'' chemical environments to give the model high accuracy where needed, as well as sufficient constraints for it to be valid. Once representative structures (in ML terms, {\em data locations}) have been chosen, they are given reference values ({\em data labels}): energies, forces, and stresses as computed with the ground-truth method. For inorganic materials, the latter is typically some flavor of density-functional theory (DFT). Higher-level, beyond-DFT computations are beginning to be used as well. \cite{Liu2022} 

In terms of how reference data determine the quality of an ML potential (Fig.\ \ref{fig:MLP_overview}b), two aspects are relevant here -- corresponding to the horizontal and vertical axes in the sketches of Fig.\ \ref{fig:MLP_overview}, respectively. On the one hand, the judicious choice of data {\em locations} (for which configurations exactly do we want to fit?) is important: this choice can be guided by the practitioner's physical and chemical knowledge, or by automated active-learning approaches for steering the database building, \cite{Podryabinkin2017, Zhang2018, Kozinsky2020} or both. However, even with carefully crafted datasets, there remains a risk of potentials not having ``seen'' what they need, resulting in poor extrapolation outside the training domain and consequently in incorrect behavior.
On the other hand, the nature and quality of the data {\em labels} is important. Any ML potential will reproduce at best the level of data on which it has been trained: for example, if the reference data labels have been computed with a simple DFT functional that does not correctly capture van der Waals interactions, neither will the resulting ML potential. The role of the numerical quality of the data (e.g., the fact that strict convergence with $k$-point sampling is required) has been pointed out in Ref.\ \citenum{Deringer2021a}, and has been systematically analyzed recently. \cite{Bayerl2022}

\begin{figure*}
    \centering
    \includegraphics[width=17.8cm]{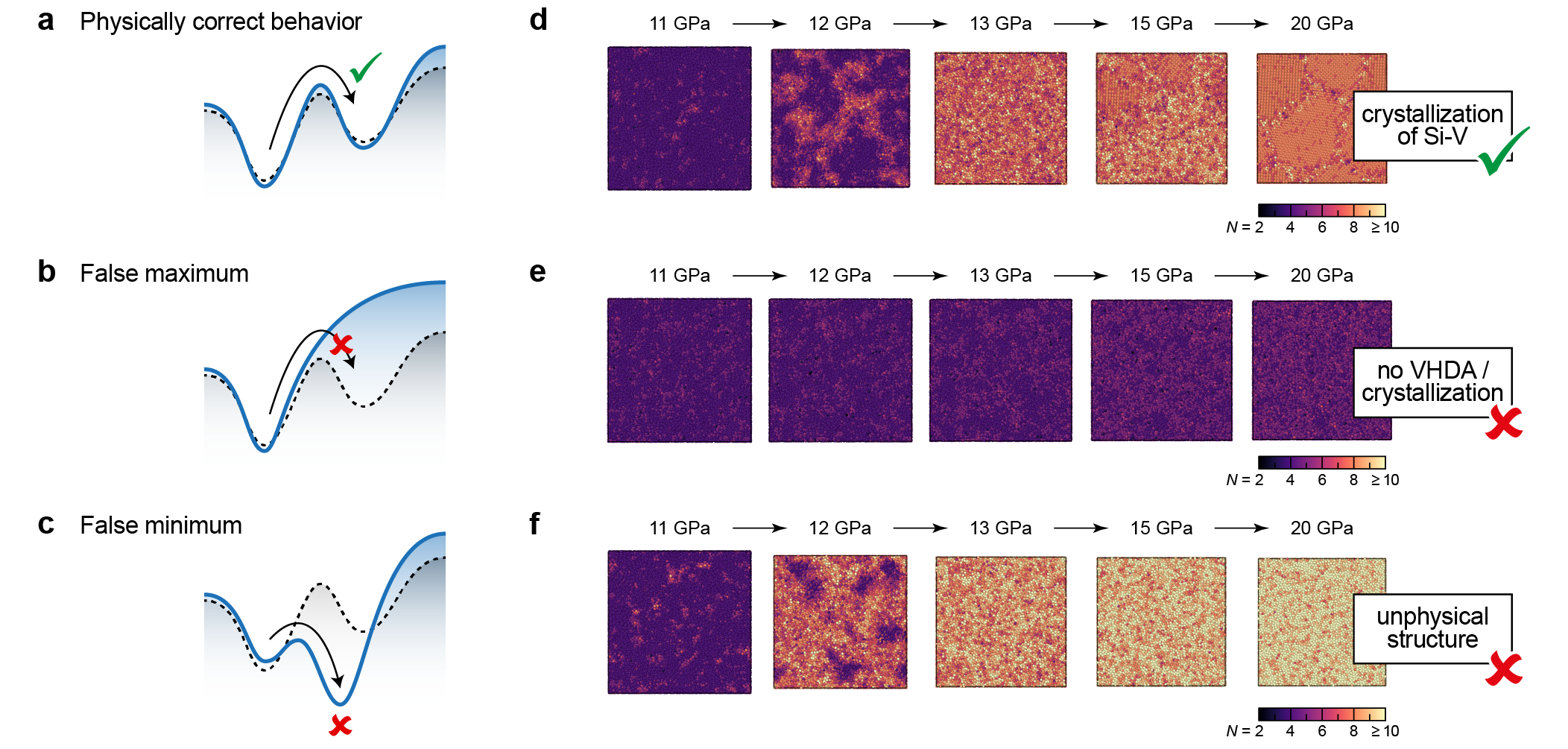}
    \caption{Validity of interatomic potentials and the effect of false maxima or minima.
    (\textbf{a}--\textbf{c}) Schematic drawings indicating the fit between a reference potential-energy surface (PES), sketched as dashed lines, and the potential model that approximates it, sketched as solid blue lines. If a false maximum exists, the simulation is not able to reach the relevant local minimum. If a false minimum exists, the potential will readily visit a spurious minimum that corresponds to a physically incorrect structure, and once there the simulation cannot easily revert to other regions of the PES (even if those were correctly described).
    Drawn following Ref.\ \citenum{Behler2021}.
    (\textbf{d}--\textbf{f}) Examples from compression simulations for silicon. The reference is a recent study of the high-pressure structural changes. \cite{Deringer2021} 
    The ``false maximum'' scenario is indicated by a simulation of the same type with an empirical interatomic potential, which does not access the very-high-density amorphous (VHDA) phase. The ``false minimum'' scenario is exemplified by a candidate potential model which shows spurious formation of an unphysical highly-coordinated structure, and was therefore identified as an non-valid candidate.\cite{Morrow2022} Panels (d) and (e) were drawn with data from Ref.\ \citenum{Deringer2021}, panel (f) with data from Ref.\ \citenum{Morrow2022}.
    }
    \label{fig:Si_compression}
\end{figure*}

The second step is to represent the local environments of the atoms in the database in a mathematical form that is suitable for learning. The tool for this task is most commonly called a {\em descriptor} in the community, and is analogous to a set of {\em features} in ML research. A good descriptor is invariant (unchanged) with respect to translations, rotations, and permutations of atoms, and is as complete as possible \cite{Pozdnyakov2020} whilst remaining numerically efficient. Many currently established ML potentials rely on hand-crafted descriptors constructed from radial and angular basis functions, \cite{Behler2011, Bartok2013, Thompson2015} and other recent approaches ``learn'' a structural representation as part of the potential fit. \cite{Schuett2018, Batzner2022} The construction of structural descriptors for atomistic ML has been reviewed in Ref.\ \citenum{Musil2021}.

Again, the user's choices here will affect the quality of the resulting potential (central panels in Fig.\ \ref{fig:MLP_overview}b). For example, models based purely on pair-wise, ``2-body'' descriptors may work well for bulk metals, but not for covalent systems. \cite{Glielmo2018} Then there is the choice of {\em hyperparameters}, i.e., of those parameters that are not directly optimized when training a single instance of a model. As one of many examples, the cartoon in Fig.\ \ref{fig:MLP_overview}b indicates the varied atomic-density broadening that can be chosen in the Smooth Overlap of Atomic Positions (SOAP) descriptor: \cite{Bartok2013} larger values can make potentials robust enough for structure searching, whilst smaller values are needed for highly accurate predictions. \cite{Deringer2018} 

The third step is to fit a flexible (highly-parameterized) function to the reference data. Among the main classes of fitting approaches currently used for ML potentials, there are: (i) artificial neural network (NN) models such as Behler--Parrinello-type NNs \cite{Behler2007, Artrith2016, Smith2017} or the DeepMD,\cite{Zhang2018a} SchNet, \cite{Schuett2018} and NequIP \cite{Batzner2022} schemes; (ii) kernel-based methods such as the Gaussian Approximation Potential (GAP) framework; \cite{Bartok2010} and (iii) linear models including the Spectral Neighbor Analysis Potential (SNAP), \cite{Thompson2015} Moment Tensor Potential (MTP), \cite{Shapeev2016} and Atomic Cluster Expansion (ACE) \cite{Drautz2019} techniques. There are interesting connections between the different methodologies -- for example, a recently proposed ``multi-layer'' ACE approach \cite{Bochkarev2022} combines the ACE descriptors with message-passing NN architectures. We leave details of the available fitting methodologies to recent review articles. \cite{Unke2021, Behler2021a, Deringer2021a}

The third set of user choices therefore concerns the regression framework itself.\cite{Pinheiro2021} Does one want a highly flexible deep-learning model requiring lots of data, or a ``tailored'' kernel-based approach that can make do with fewer examples? What are the hyperparameters of the fit: in GAPs, the regularization (``expected error'') is important;\cite{Deringer2021a} NNs depend on the learning rate and batch size; other methods will be affected by other aspects. For the purpose of the present paper and the following discussion, any ML potential fitting framework will be relevant in equal parts.

\section{What makes a potential ``valid''?}

Overall, the answer seems to be easy. A new ML potential, and indeed any interatomic potential model, has to pass two tests. Does it {\em qualitatively} predict what it should (to the extent that is comparable with experiment)? And does it {\em quantitatively} do so -- say, in predicting measurable properties? 

In practice, it is often very difficult to determine the quality of a potential, and to connect the numerical performance to its physical meaning. Therefore, validation tests become important. Figure \ref{fig:Si_compression} summarizes a way to think about the qualitative ``correctness'' of a potential in terms of false maxima and minima (and their effect on predictions), respectively. This issue was discussed in recent review and perspective articles. \cite{Deringer2021a, Behler2021} 

Qualitatively correct potentials are expected to show the correct behavior for a given system (with respect to either experiment, or a high-quality quantum-mechanical prediction, or both): we illustrate this on the right-hand side of Fig.\ \ref{fig:Si_compression}d, which shows snapshots from the simulated compression of amorphous silicon, as taken from recent work. \cite{Deringer2021} 
The key observation in that study -- and thus the benchmark that a ``correct'' simulation will need to reproduce -- was the pressure-induced crystallization to form simple-hexagonal silicon (Si-V). The coordination number in that phase is eight, and so the final poly-crystalline structure is rendered in orange in Fig.\ \ref{fig:Si_compression}d.

The figure then compares the ``correct'' prediction (Fig.\ \ref{fig:Si_compression}d) to the same simulation with a more limited, yet much faster, empirical interatomic potential (Fig.\ \ref{fig:Si_compression}e), and finally to the result of a candidate ML potential that is deliberately chosen as an example of one that is {\em not} correct (Fig.\ \ref{fig:Si_compression}f). In the former case, which we take to illustrate a false maximum, the structure remains relatively similar to the low-coordinated low-density form of amorphous silicon (purple) -- no structural collapse nor crystallization happens.\cite{Deringer2021} We presume that this is because the functional form of the empirical potential  used is not as flexible as that of a typical ML potential, instead favoring the tetrahedral local geometry inherent to low-density silicon. 
For the latter case (Fig.\ \ref{fig:Si_compression}f), we show a candidate ML potential that predicts an unphysical structure: ``unphysical'' is here taken to mean that the system does not crystallize, rather getting stuck in a fully disordered configuration with coordination numbers of $\geq 10$ (Ref.\ \citenum{Morrow2022}).

A potential which is not qualitatively valid (such as the one in Fig.\ \ref{fig:Si_compression}f) should not be used for simulations. A potential which is qualitatively valid next needs to be assessed with regards to its {\em quantitative}, numerical accuracy. This will be the topic of the following section.

\section{Numerical errors}

Numerical error metrics are of central importance in many areas of ML -- for example, in quantifying the performance of a new model compared to the existing state of the art (SOTA). 
We briefly review relevant techniques for numerical validation and error definitions that are commonly used in the ML literature, and then discuss their application to interatomic potential models.

\subsection{Error measures}\label{sec:error_measures}

\begin{figure}
    \centering
    \includegraphics[width=8.5cm]{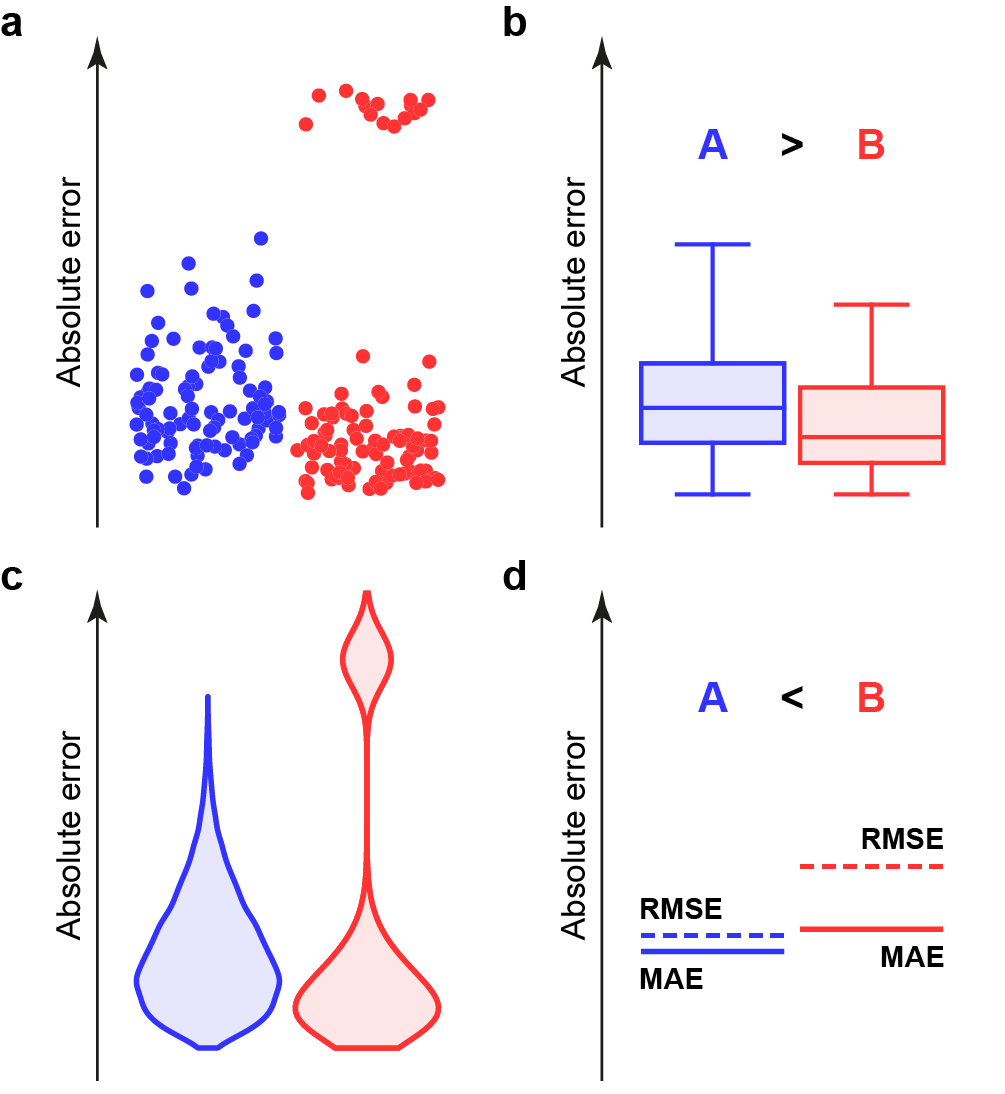}
    \caption{Numerical errors and their visualization. The figure compares various ways to graphically explore error for two hypothetical ML potentials, \textbf{A} on the left and in blue, and \textbf{B} on the right and in red: (\textbf{a}) data points generated using a random number generator, in arbitrary order, (\textbf{b}) box and (\textbf{c}) violin plots, together with (\textbf{d}) error metrics on the same axes (absolute error). 
    Graphical methods provide information as to the \textit{distribution} of errors, as opposed to error metrics which summarize this information into a single measure. 
    As discussed in the text, we argue that model \textbf{B} is worse than model \textbf{A} for the purpose of materials modeling, due to its large departure from the true PES in some places. However, if non-robust metrics such as median and percentile errors were used to quantify performance (cf.\ the median values shown as horizontal lines in panel b), they would suggest the opposite.
    }
    \label{fig:error-dist}
\end{figure}

When testing any ML regression model, a predicted label, $\hat{y}_i$, is generated for each entry (each data location) in a test set of data that have not been included in training the model.
Each predicted label corresponds to a correct ground-truth value, $y_i$, and numerical error metrics are designed to summarize, in a single measure, the disparity between prediction and ground truth as taken over all data points in the test set. Different error metrics capture different qualities of this disparity.

\begin{figure*}
    \centering
    \includegraphics[width=17.8cm]{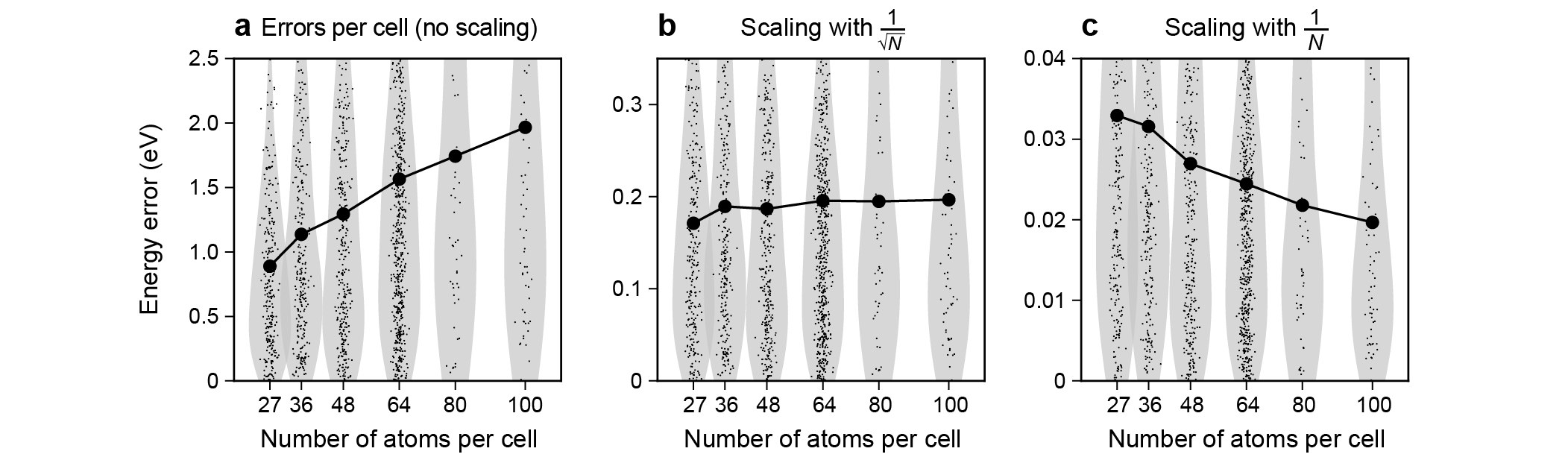}
    \caption{
    Error scaling with system size. Moving from left to right, we plot errors for total structure energy predicted by an existing ML potential, \cite{Deringer2017} as normalized using: no scaling factor, scaling of $1/\sqrt{N}$, and the commonly-used scaling of $1/N$.
    Raw data points, i.e., errors for individual structures, are plotted using small symbols, with their corresponding distributions represented in the light gray violin plots. Mean errors for each structure size are plotted on top of these.
    Both the violin and mean plots show that a scaling of $1/\sqrt{N}$ leads to a distribution of normalized errors that is independent of system size. The same is not true for either of the other cases.
    Errors were evaluated for predictions made by the GAP-17 ML potential (as a sum of local energies) compared to ground-truth DFT total energies. A subset of the full GAP-17 reference database (Ref.\ \citenum{Deringer2017}) was selected for this exercise, such that the type and energy distribution of structures is the same for all $N$.
    }
    \label{fig:scaling}
\end{figure*}

Two of the most commonly used metrics are the mean absolute error (MAE) and the root mean square error (RMSE), defined respectively as
$$
\textrm{MAE} = \frac{1}{N} \sum_i{\left|y_i - \hat{y}_i\right|} \qquad \textrm{RMSE} = \sqrt{\frac{1}{N} \sum_i \left(y_i - \hat{y}_i \right)^2 }.
$$
While these metrics both attempt to measure the average error, the RMSE is strictly $\ge$ MAE, and is skewed upwards by higher errors.

In many settings, one prefers robust statistics, i.e., those that are not skewed by extreme values. 
However, we argue that robust statistics should not be used {\em on their own} in validating ML potentials, based on the following scenario (Fig.\ \ref{fig:error-dist}a):
consider a potential ({\bf A}) that performs reasonably well over all relevant configurational space, and compare it to another potential ({\bf B}) that performs extremely well on 90\% of the data points, but poorly on the remaining 10\%. 
A robust error metric such as the median would identify {\bf B} as being ``better'' than {\bf A}. This is illustrated in Fig.\ \ref{fig:error-dist}b, where the middle horizontal line in the box plots indicates the median value.
In computational practice, however, a potential that fails dramatically to capture the true nature of the PES in a physically relevant region ({\bf B}) is not valid, and is therefore significantly worse than one which performs reasonably over all of configurational space ({\bf A}).

The metrics presented above summarize the distribution of residual errors using a single value. 
While offering a practical way to compare different models, they necessarily lose information about the overall distribution of errors.
Using a kernel density estimation or ``violin'' plot is a graphically succinct way to more completely describe the error distribution (Fig.\ \ref{fig:error-dist}c), and in the context of ML potentials seems preferable to a box plot (Fig.\ \ref{fig:error-dist}b) which incorporates robust statistics.
Note how in this numerical example, both the violin plot (Fig.\ \ref{fig:error-dist}c) and the individual RMSE and MAE values (Fig.\ \ref{fig:error-dist}d) provide an indication of unfavorable errors in \textbf{B} compared to \textbf{A}.

\begin{figure*}
    \centering
    \includegraphics[width=17.8cm]{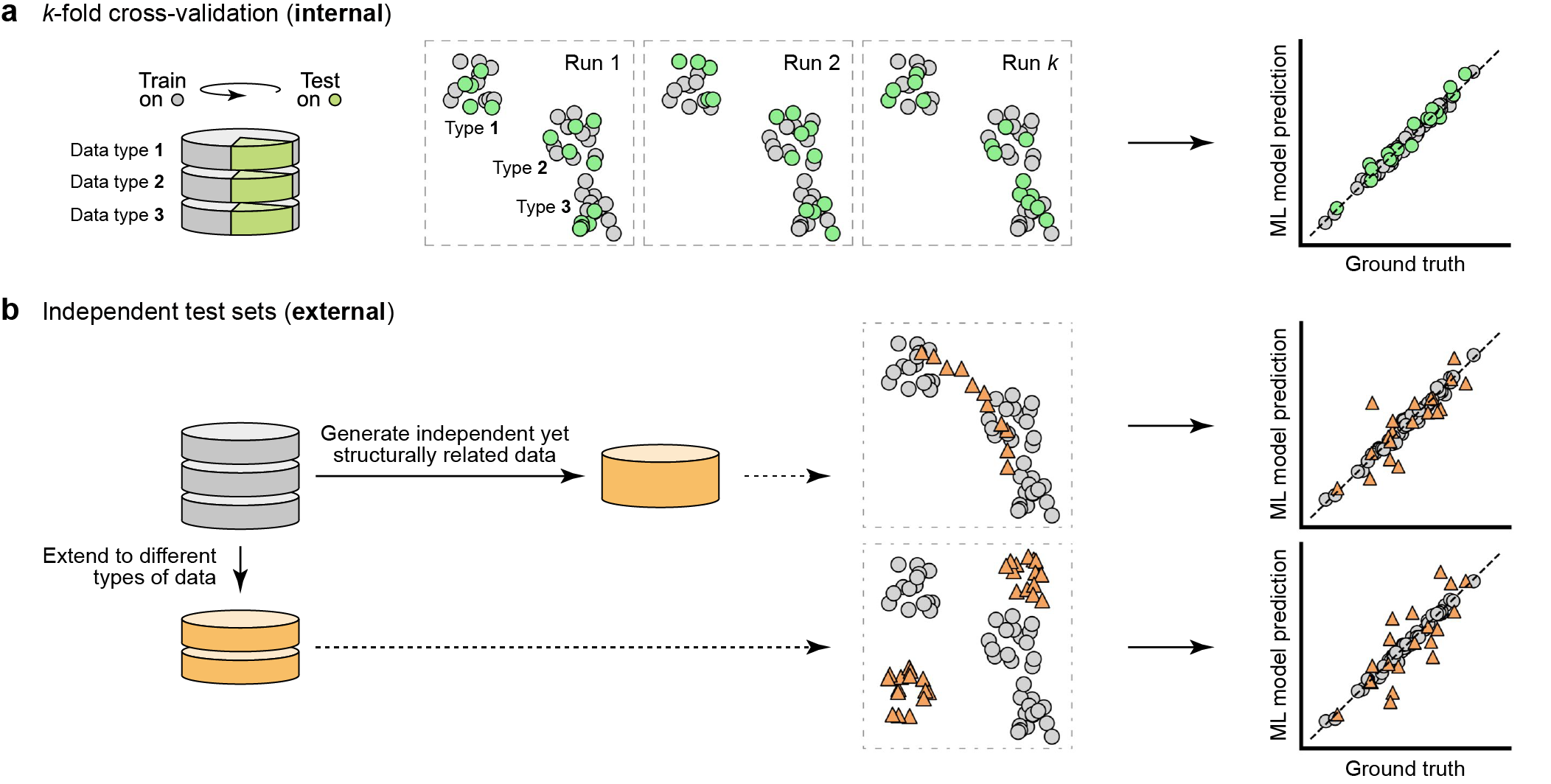}
    \caption{
    Schematic cartoons depicting two approaches to creating test sets for numerical validation, as are commonly used in ML research and in the development of ML potentials.
    The central panel of each section illustrates the location of various training and testing data sets in a dimensionally reduced space; the right illustrates a parity plot comparing the ML model predictions against the ground-truth data.
    (\textbf{a}) \textit{k}-fold (``internal'') cross-validation: the entire reference database is used to test the model, where \textit{k} successive folds are held out during training.
    (\textbf{b}) Independent (``external'') test sets: a separate data set is created explicitly for testing. Typically, this test set this derives from a different distribution to that used to train the model.
    }
    \label{fig:numerical}
\end{figure*}

\subsection{Scaling with system size}

ML potentials typically make atom-wise predictions: atomic energies (that are then summed up to give the total energy\cite{Behler2007, Bartok2010}) and forces on atoms. When predicting per-atom properties such as the magnitude or Cartesian components of forces, where there is a one-to-one correspondence between DFT ground truth and ML-predicted values, no further considerations are therefore required when reporting errors.
However, for {\em energies}, DFT does not normally provide per-atom values, and the ground truth in this case is the total energy for a given simulation cell. 
When predicting system-wide properties such as the total energy, $N$ individual ML model predictions are therefore summed and compared to a single true label. In these cases, the system size has an effect on the behavior of the numerical errors. 

Figure \ref{fig:scaling} illustrates this point using a practical example: a set of structures from the reference database of the GAP-17 ML potential for carbon, which is published alongside the original work described in Ref.\ \citenum{Deringer2017}. We evaluated the model errors for different system sizes and investigated if, and how, they change with $N$. The prediction errors for the entire cells trivially increase with $N$ (Fig.\ \ref{fig:scaling}a). In contrast, when the energy error is reported as a per-atom value (by dividing the predicted total energy by $N$), the result systematically {\em decreases} with $N$ (Fig.\ \ref{fig:scaling}c). Figure \ref{fig:scaling}b illustrates that, in line with the central limit theorem, scaling the predictions by $\sqrt{N}$ would alleviate this problem.

We therefore recommend that, when comparing model performance for per-cell properties such as the total energy, datasets containing structures of the same size are used. If this is not possible, it would seem desirable to use the statistically justified normalization procedure of dividing by $\sqrt{N}$. We do note, in practical terms, that per-cell energies normalized in this way have different absolute values from the commonly quoted error values in eV per atom (which one would obtain by dividing by $N$), and are therefore not directly comparable -- as seen from the different $y$-axis values in the panels of Fig.\ \ref{fig:scaling}.  

The most important message of this figure, rather than specific error values for a specific potential, is that numerical errors do need to be evaluated with care. Errors for a given ML potential, if quoted in isolation, will likely be of limited use; however, the evolution of a well-defined error measure across different types of fitting methods, database compositions, hyperparameters, etc., and the use of such measures in systematic benchmarks, will be (and will continue to be) highly informative.

\subsection{Cross-validation and external test sets}\label{sec:test_sets}

Many ML model classes can nearly perfectly fit to the data on which they have been trained. To measure a model's true capabilities, and in particular its ability to generalize, it is therefore important to test on data points that have not been included in the training.

A popular way to numerically validate an ML model is to use $k$-fold cross validation (Fig.\ \ref{fig:numerical}a).
This procedure involves separating the complete dataset into $k$ non-overlapping sets (or ``folds''), and training $k$ separate models, each using the remaining data when the $k$-th set is held out for testing. 
Averaging an error metric over all $k$ folds yields a measure that is less affected by random noise induced by the exact choice of train-/test-set splitting.
In Fig.\ \ref{fig:numerical}a, we sketch this process schematically for a database that consists of three clusters of data. For an ML potential, these could correspond, say, to bulk crystalline, liquid, and surface structural models, respectively -- see Ref.\ \citenum{Rowe2020} for an example of how such a dataset might look in practice. The sketches in dashed boxes indicate the distribution of the data points in a 2D projection of structural similarity (similar structures being close together, and vice versa), with the three types of data forming three clusters, and different points being used as test data (green) in each of the $k$ folds. The prediction of the ML model is then plotted against the ground-truth value for every point as measured when it was used for testing, and error metrics such as the RMSE (Sec.\ \ref{sec:error_measures}) can be calculated.

An alternative, more time-consuming, procedure for numerical validation involves generating one or more ``external'' test sets. These test sets could be generated in similar ways as the training data, that is, be independent yet structurally related -- or they could extend to different types of data (Fig.\ \ref{fig:numerical}b).

In atomistic materials modeling, an example of a general, system-agnostic approach is a standardized random structure search (RSS) protocol; \cite{Pickard2006, Pickard2011, Bernstein2019b} these searches explore a wide range of configurations, thus generating an unbiased sample of diverse atomic structures, including local minima in the PES. We discuss RSS-based test sets in more detail in Sec.\ \ref{sec:RSS} below.
The creation of more specific test sets can be guided using domain knowledge about the chemical system -- for instance, surfaces, defects, transition paths, and polymorphs not seen during training.
To more fully understand behavior in a particular domain, these sets can be highly specialized, for instance containing solely a set of manually distorted crystal structures, or snapshots from an MD simulation at high temperature and pressure.
The error values obtained using this technique will depend strongly on the nature of the test set, and baselining using several different test sets offers a way to more comprehensively understand model behavior.

We note that our wording ``external'' in Fig.\ \ref{fig:numerical}b refers to the construction process of the test set, rather than to its location relative to the training set. Comparing the two test sets sketched in Fig.\ \ref{fig:numerical}b, data located in regions close to training points (upper panel) will typically be modeled more accurately than those further away (lower panel). Such ``training example data leakage'' can lead to overly optimistic error estimates, unless the test set accurately reflects the data to which the model will be applied.

\section{Physically-guided validation}

In this section, we discuss a series of physically-guided tests for ML potentials for materials. This area is one where the validation becomes very different from techniques used in ``standard'' ML research, and relies on the practitioner's domain knowledge. We argue that such physically-guided validation plays an important role in making an ML potential applicable in practice.

\subsection{Domain-specific error analysis}\label{sec:domain_specific_errors}

The first of our proposed ``physically-guided'' tests is actually still to do with energy and force errors. The key difference is that we generate external testing data entirely separately, in a setting which is informed by physical and chemical knowledge as far as possible.

\begin{figure}
    \centering
    \includegraphics[width=8.5cm]{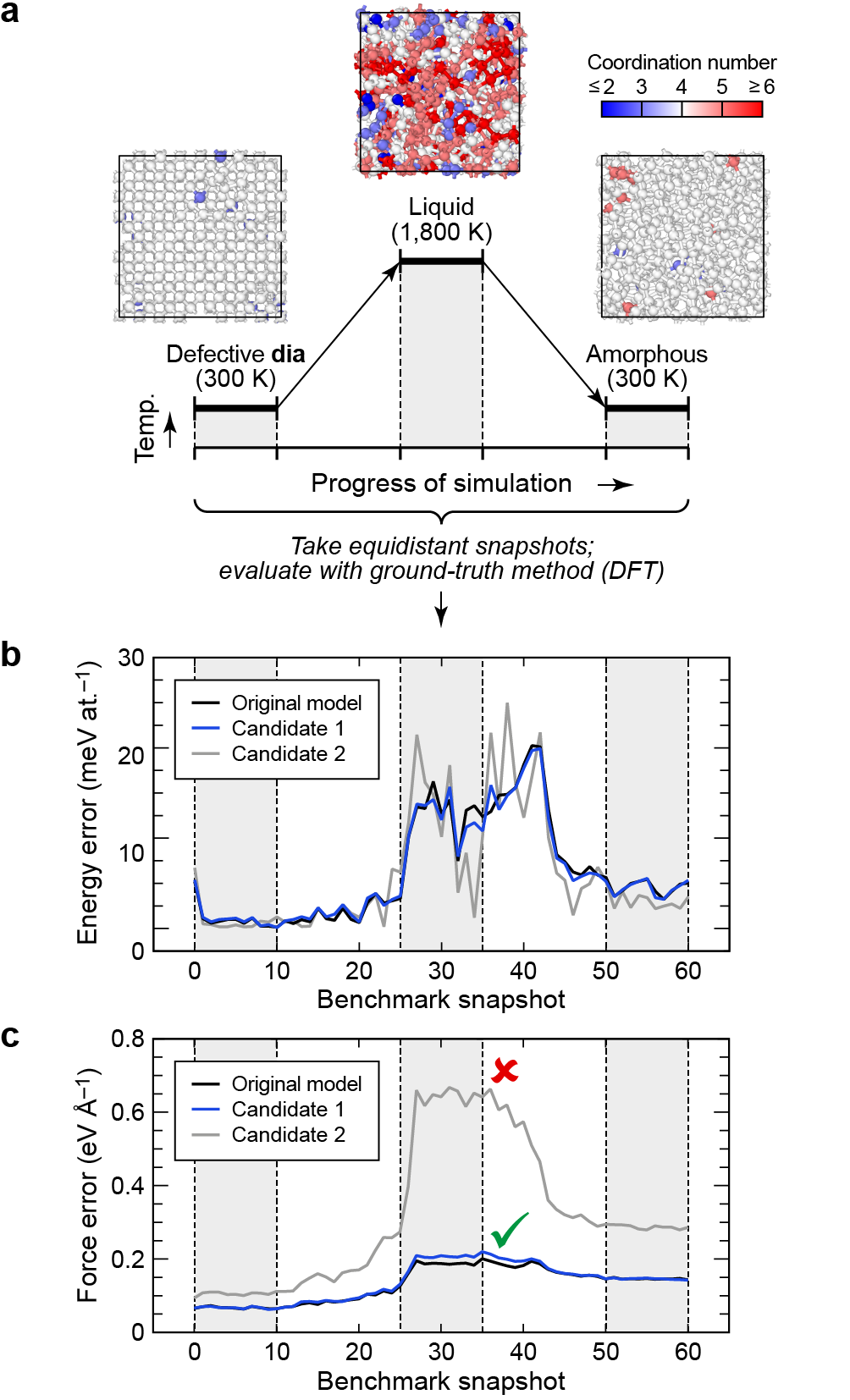}
    \caption{A physically-motivated DFT benchmark for silicon ML potentials. \cite{George2020}
    (\textbf{a}) Schematic of an ML-driven melt--quench simulation that is designed to encompass a range of relevant scenarios: the description of a highly defective crystalline phase as well as melting and quenching. The simulation proceeds over a total of 600 ps for a system of 500 atoms.
    (\textbf{b}) Energy errors for 61 snapshots along the trajectory, evaluated for the original model (GAP-18, from Ref.\ \citenum{Bartok2018}, taken as the current state-of-the-art) and two candidate potentials. As discussed in the original work, \cite{George2020} these two potentials are optimized for accurate phonon predictions, and the test here is designed to see whether they still remain valid for a different application case.
    (\textbf{c}) Same but for errors of the Cartesian force components (which are important to distinguish from force magnitudes or directions).
    This plot reveals how the first candidate potential performs well (blue line), whereas the second fails severely for the description of disordered silicon (gray line).
    Adapted from Ref.\ \citenum{George2020}.
    Original work published under a Creative Commons Attribution license (http://creativecommons.org/licenses/by/4.0/).
    }
    \label{fig:Si_JG}
\end{figure}

In Fig.\ \ref{fig:Si_JG}, we show the construction and use of such a physically motivated test set for ML potentials, taken from original work in Ref.\ \citenum{George2020}. The idea is to carry out a small-scale molecular-dynamics (MD) simulation that mimics the ``real thing'' -- small enough so that structural snapshots from that MD trajectory can be evaluated (labeled) in subsequent single-point computations with the ground-truth method. In this case, the reference MD simulation was run with an ML potential (referred to as ``Original model'' in Fig.\ \ref{fig:Si_JG}), and an important prerequisite for doing so was that that potential had itself been validated in earlier work. \cite{Bartok2018, Deringer2018d} In the study of Ref.\ \citenum{George2020}, the aim was to test the behavior of modified GAP models for elemental silicon, which were based on the original GAP-18 potential \cite{Bartok2018} and included additional data and custom regularization to more accurately describe diverse crystalline allotropes and their vibrational properties. \cite{George2020}

\begin{figure*}
    \centering
    \includegraphics[width=17.8cm]{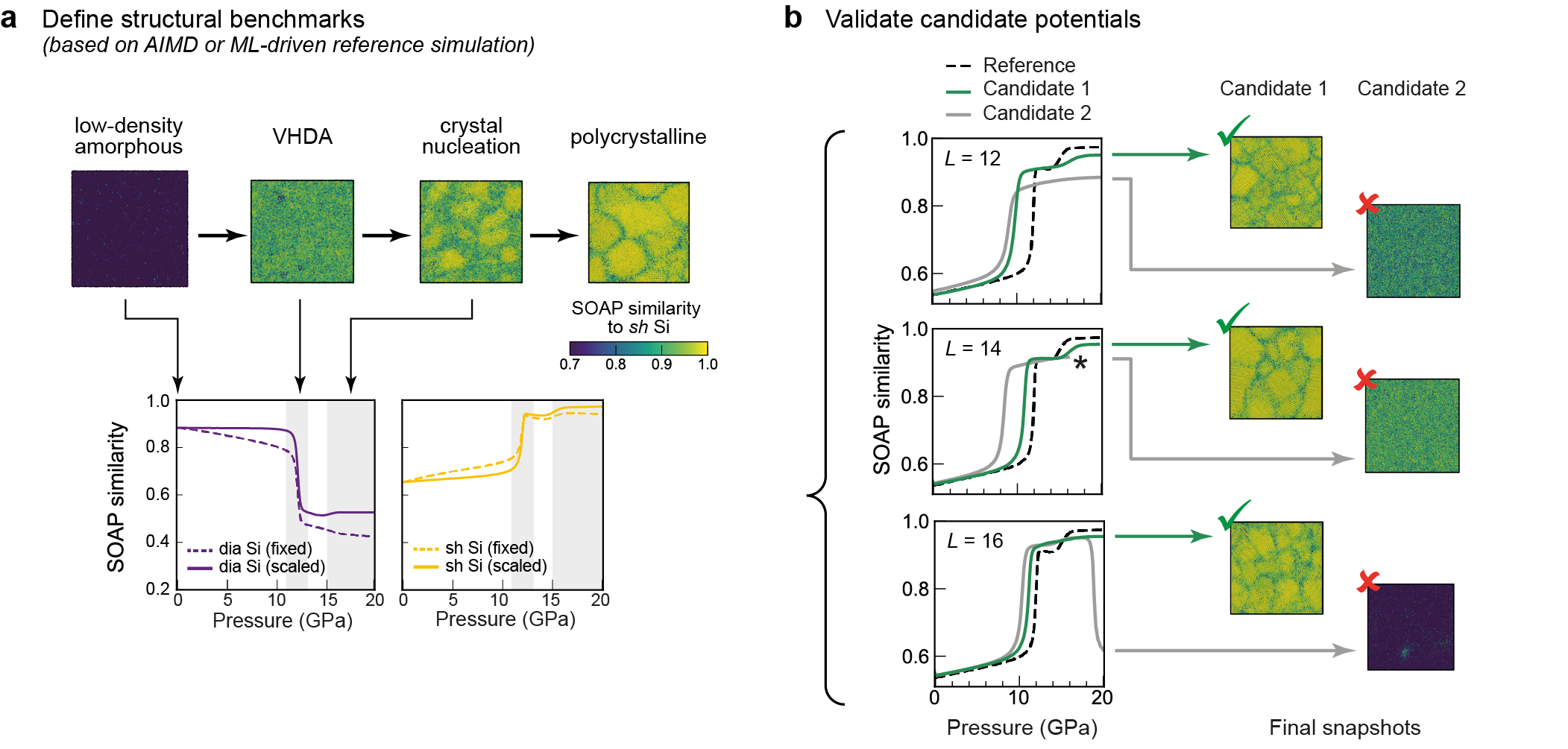}
    \caption{Structural-similarity-based validation against a reference simulation, using the SOAP kernel to quantify similarity. (\textbf{a}) Snapshots from the compression simulation of Ref.\ \citenum{Deringer2021}, taken at 0, 12, 15, and 20 GPa, respectively. Atoms are colored according to their SOAP similarity to crystalline simple hexagonal (sh) Si. The panels below show the average SOAP similarity of the compression trajectory to (left) diamond-type and (right) sh silicon during compression. Dashed lines correspond to comparisons to a fixed reference crystal at ambient pressure, solid lines to a crystal with an external stress that is equivalent to the pressure at each snapshot during the MD trajectory. (\textbf{b}) As before, but now comparing two types of candidate potentials, each row with different hyperparameter settings, as used to perform an MD simulation of the same protocol as panel (a). Comparisons of SOAP similarity to sh silicon reveal qualitatively incorrect trajectories predicted by candidate 2, including unstable MD marked by an asterisk, and acceptable predictions from candidate 1. Adapted from Ref.\ \citenum{Morrow2022}. Original work published under a Creative Commons Attribution license (http://creativecommons.org/licenses/by/4.0/).}
    \label{fig:soap_validation}
\end{figure*}

In this example, two new candidate ML potentials are tested, which only differ in one aspect of the fit -- namely, in the choice of the regularization hyperparameters corresponding to the ``expected error'' in the input data. (In the schematic in Fig.\ \ref{fig:MLP_overview}b, all aspects would therefore be the same except for the last one in the series.) Candidate potential 1 is a model where new structures have been added to the existing GAP-18 database, but not all too much weight is placed on them. Candidate 2 is a model where the regularization is ``tighter'' by a factor of 10, and so the potential is very good indeed at describing phonons, but at the expense of physically reasonable behavior outside that scope. In particular, an MD melt-quench simulation using candidate 1 reproduced the behavior of the original GAP-18 model, whereas the same simulation using candidate 2 failed entirely. \cite{George2020}

Force errors on a physically motivated test set can be predictive of this behavior, as evidenced in Fig.\ \ref{fig:Si_JG}c. By contrast, we emphasize that the energy errors on their own do not reveal any trouble (Fig.\ \ref{fig:Si_JG}b), other than a slightly larger scatter compared to the original GAP-18 model. In fact, candidate 2 was found to have a {\em lower} numerical energy error for the liquid configurations in this test set (8.4 meV/atom) than candidate 1 (11.5 meV/atom). \cite{George2020}

\begin{figure*}
    \centering
    \includegraphics[width=17.8cm]{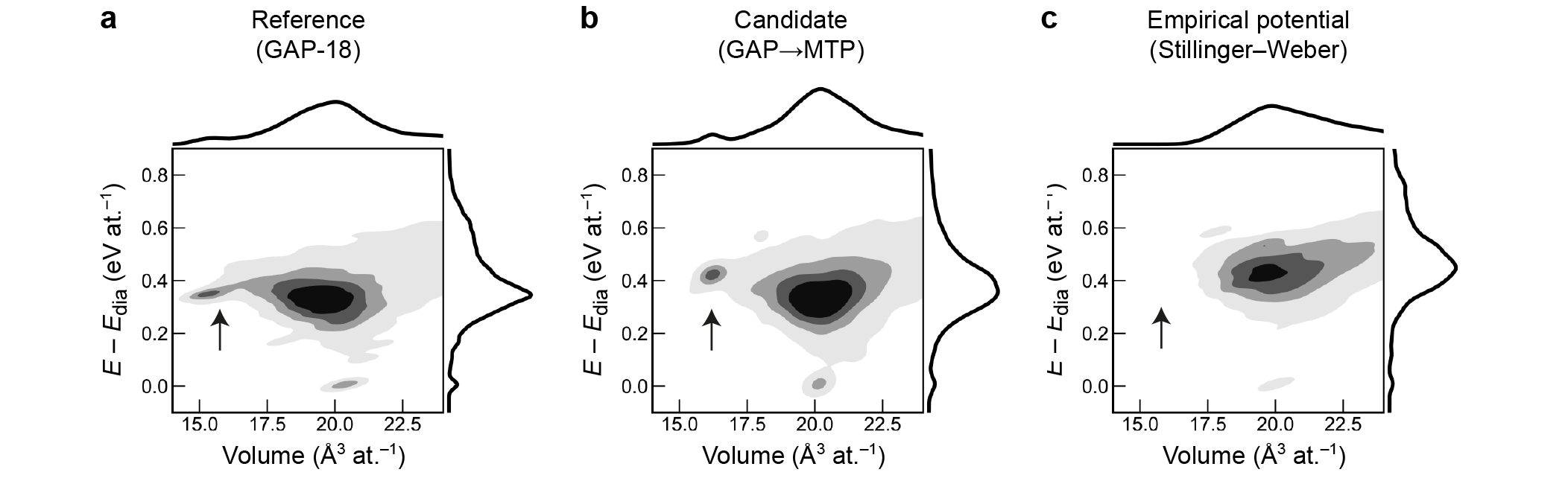}
    \caption{Random structure search (RSS) for silicon with 10,000 structures with (\textbf{a}) the GAP-18 model from Ref. \citenum{Bartok2018}, (\textbf{b}) the ``candidate 1'' $L=16$ MTP model characterized in Fig.\ \ref{fig:soap_validation} (from Ref.\ \citenum{Morrow2022}), and (\textbf{c}) the empirical Stillinger--Weber potential. \cite{Stillinger1985} The contours represent the density of points, and the histograms show their corresponding distributions in energy and volume, respectively. Arrows highlight the presence or absence of RSS minima similar to the high-pressure forms of silicon.
    This test is inspired by Ref.\ \citenum{Bartok2018}; however, all searches are newly conducted in the present study.
    }
    \label{fig:rss}
\end{figure*}

\subsection{Domain-specific structural benchmarks}

The second type of physically-motivated tests concerns structural similarity analysis, which we carry out using the SOAP kernel. \cite{Bartok2013} As in the previous section, we suggest to validate a candidate potential by comparing its behavior to an accurate reference simulation, which could be based on DFT (small scale), or driven by an existing and previously validated ML potential (large scale). \cite{Morrow2022}

As an illustrative example, similar to Fig.\ \ref{fig:Si_compression}, we use as benchmark the results of large-scale (100,000-atom) MD simulations of the pressure-induced crystallization of amorphous silicon. \cite{Deringer2021} In Fig.\ \ref{fig:soap_validation}a, we color selected structural snapshots by the atomistic SOAP kernel similarity to the crystalline simple hexagonal (sh) phase. This color-coding clearly highlights the two significant structural changes: collapse of the fourfold-coordinated amorphous phase, followed by nucleation of sh crystallites. 
Figure \ref{fig:soap_validation}a shows the average SOAP similarity, taken over the whole simulation cell at each timestep. The interpretation of this similarity measure is intuitive: the atomic environments in the low-density amorphous phase are mainly tetrahedral-like, and similar to the diamond-type crystalline form (high SOAP similarity value), then become dramatically less diamond-like upon structural collapse. The similarity to sh silicon follows the opposite trend, with the very high similarity at the end of the simulation positively identifying the sh crystallites in comparison to other possible competing phases (see Ref.\ \citenum{Morrow2022} for more details). 

There is a subtle detail in constructing these similarity plots. For the solid lines in Fig.\ \ref{fig:soap_validation}a, we relax the reference crystal structure under an external pressure that matches that of the corresponding frame in the MD simulation. This approach accounts for the change in bond lengths with pressure, which is most evident in the SOAP similarity of the low-density amorphous phase to diamond at 0 to 10 GPa. Even though no significant structural rearrangement occurs, the similarity to the fixed reference crystal at ambient pressure decreases linearly (dashed line) -- in contrast, if a pressure-adjusted crystal is used as reference, almost no change in the SOAP similarity is seen up to about 10 GPa (solid line).

Once a structural-similarity benchmark has been developed, it can be used to assess new candidate ML potentials. In the case we review here, originally reported in Ref.\ \citenum{Morrow2022}, the aim was to train computationally much cheaper potentials that still show the same physical behavior as their ``teacher'' model. In Fig.\ \ref{fig:soap_validation}b, we use the quantitative structural metric provided by SOAP to compare predictions of candidate ``student'' potentials to the previously-validated reference simulation of Ref.\ \citenum{Deringer2021}.
Specifically, we generated two sets of MTP models which are controlled by their maximum level, $L$, fitted separately to a large database of structures (candidate 1) and to a comparatively small one (candidate 2). Unreasonable, highly-coordinated false minima (of differing kinds) are detected by the SOAP analysis for candidate 2. Instead, candidate 1 passed the test and could therefore be used for large-scale MD simulations with confidence. \cite{Morrow2022}

More generally, we think that compression MD simulations starting from some highly disordered structure can provide an insightful test for the physical behavior of ML potentials -- particularly if a large system size is used, allowing for the frequent sampling of a range of configurations involving the close approach of atoms. The increasing comprehensiveness of easily-accessible structural databases, such as the Materials Project, \cite{Jain2013} means that reference crystal-structure data are readily available for many chemical systems, and structural-similarity analyses such as that exemplified in Fig.\ \ref{fig:soap_validation} can be set up easily.

\subsection{Random search and exploration}\label{sec:RSS}

Random searching was first introduced as an approach to first-principles crystal-structure prediction, in the {\em Ab Initio} Random Structure Searching (AIRSS) framework by Pickard and Needs.\cite{Pickard2006, Pickard2011} AIRSS aims to discover previously unknown crystal structures by generating and relaxing (with DFT) large numbers of random structures that satisfy some simple constraints, such as the minimum separation between atoms,\cite{Pickard2011} and a similar approach can be taken with ML potentials. \cite{Deringer2018, Pickard2022}
The task of relaxing random structures into their local, often high-energy, minima provides a stringent test for an interatomic potential. This type of test has been introduced for silicon, where a range of widely used empirical potentials do not reproduce the energy distribution of the RSS minima as closely as an ML potential.\cite{Bartok2018}

\begin{table*}[t]
    \caption{Experimental observables against which predictions from an ML potential may be compared. Dots indicate the, somewhat subjectively evaluated, difficulty of a given experiment or computation, from easy ($\bullet$) to challenging ($\bullet\!\bullet\!\bullet$).\\[2mm]}
    \centering
    \begin{tabular}{lp{4mm}p{56mm}p{4mm}p{56mm}}
    \hline
    \hline
        \textbf{Quantity} & & \textbf{Experimental technique} & & \textbf{Computational counterpart} \\
    \hline
        Lattice parameters & & X-ray ($\bullet$) or neutron diffraction ($\bullet\!\bullet\!\bullet$)  & & Structural relaxation ($\bullet$)  \\
    \hline
        Excess enthalpy 
        & & Calorimetry ($\bullet \bullet$) 
        & & Energy ($\Delta E$, $\bullet$) or enthalpy ($\Delta H$, $\bullet \bullet$) \\
    \hline 
        Vibrational spectroscopy & & Inelastic neutron scattering   ($\bullet\!\bullet\!\bullet$) & & Vibrational density of states from MD or phonon computations ($\bullet\bullet$) \\
                                         \cline{3-5}
                                         & & Infrared or Raman spectroscopy ($\bullet$) & & As above, but with IR/Raman intensities predicted ($\bullet\!\bullet\!\bullet$) or ignored ($\bullet\bullet$) \\
    \hline
        Atomic spectroscopy & & NMR ($\bullet\bullet$), X-ray spectroscopy ($\bullet\!\bullet\!\bullet$) & & Not normally directly, but can be computed using DFT on small ML-generated structural models ($\bullet\bullet$) \\
    \hline
        Disordered structure & & Pair distribution function (PDF) analysis ($\bullet\!\bullet\!\bullet$) & & Radial distribution function from MD simulation ($\bullet$) \\
                             \cline{3-5}
                             & & X-ray or neutron structure factor, $S(q)$, for liquid and amorphous phases ($\bullet\!\bullet\!\bullet$) & & Fourier transform of radial distribution function ($\bullet$) \\
    \hline
    \hline
    \end{tabular}
    \label{tab:observables}
\end{table*}

To illustrate the use of RSS in assessing and validating interatomic potentials, we show in Fig.\ \ref{fig:rss} the energies and volumes of 10,000 random silicon structures that have been relaxed into local minima separately with: (a) the general-purpose GAP-18 model taken from Ref.\ \citenum{Bartok2018}, which had previously been shown to reproduce AIRSS results well; \cite{Bartok2018} (b) a candidate indirectly-learned potential (GAP$\rightarrow$MTP) that has been trained on GAP-18 data; \cite{Morrow2022} and (c) the empirically fitted Stillinger--Weber (SW) potential \cite{Stillinger1985} which is widely used for modeling silicon. 
For both ML potentials, a distinct basin at low volumes can be observed (arrows in Fig.\ \ref{fig:rss}), corresponding to structures similar to simple-hexagonal-like phases at high pressures.  The empirical SW potential, by contrast, fails to relax the same corresponding random structures into this chemically-sensible local minimum, because it strongly favors diamond-like, lower-density structures. (This observation is consistent with the fact that SW predicts no structural collapse under pressure; Fig.\ \ref{fig:Si_compression}e and Ref.\ \citenum{Deringer2021}.) All three potentials do find diamond-like minima, which we define by the relaxed structure having a SOAP similarity of $> 0.99$ to the ideal crystalline form; however, the empirical potential finds considerably fewer: 112 (GAP-18), 135 (MTP), 55 (SW).

In a different vein, we mention that the structures produced by an RSS run -- both the relaxed ones and the points along the minimization trajectory -- also constitute an unbiased set that can be useful for out-of-sample testing. To this end, the RSS structures can be ``labeled'' with the ground-truth method, and any candidate potential can be compared against this set. The numerical errors in this case will likely be relatively high for these rather unusual structures, underscoring the need for viewing the absolute error values in context. We have shown an example of this in recent work.\cite{Morrow2022}

\subsection{Experimental data}

Ultimately, the test for a simulation is whether it agrees with (and explains) experimental observations. Therefore, the direct validation against experimental data is perhaps the most important and relevant test for an ML potential. Often, validation of ML potentials for materials includes {\em some} comparison with previously published experimental data. There is, however, wide scope for how exactly this comparison might be made. 

We summarize relevant techniques in Table \ref{tab:observables}. 
Crystalline materials are widely characterized by X-ray and neutron diffraction experiments, yielding lattice parameters that may be compared to those for a relaxed structure. The comparison is straightforward, and yet it is important to choose the most appropriate reference data: low-temperature measurements are closer to the ``zero-Kelvin'' simulation than those at room temperature; high-resolution data from synchrotron experiments are better (but much more scarce) than data from in-house diffractometers; powder diffraction data are preferable for lattice parameters whilst single-crystal diffraction yields the most accurate atomic positions, all else being equal. The Inorganic Crystal Structure Database (ICSD; Ref.\ \citenum{Zagorac2019} and references therein) can be helpful in locating experimental data.

Calorimetric measurements give information about enthalpy ($\approx$ energy) differences between different phases, say two crystalline polymorphs of a material, or an amorphous phase compared to its crystalline counterpart. In Ref.\ \citenum{Erhard2022}, for example, when validating an ML potential for \ce{SiO2}, it was shown that a particular DFT level (the SCAN functional) gives energetics compatible with experiment, and this good performance is inherited by the ML model. We note that with regard to validation, there are two separate effects here, viz.\ the error of the ground-truth computation compared to experiment, and the error of the ML fit to the reference database. It is therefore important to disentangle both.

The vibrational properties are another useful quantity, and the phonon dispersions for crystalline phases are now easy to predict computationally -- for example, using the {\tt phonopy} software.\cite{Togo2015a} The experimental data are somewhat more difficult to come by if one is interested in the full phonon spectrum. Whilst measuring the entirety of the vibrational density of states (VDOS) requires inelastic neutron scattering techniques, the much more common infrared and Raman spectroscopy are routine characterization tools in the laboratory. In turn, the latter types of spectra are more difficult to predict computationally: they require the (rather intricate) computation of absorption intensities associated with given phonon modes. ML models have begun to be developed for this purpose, \cite{Gastegger2017} and it is likely that ``integrated'' ML models that combine electronic predictions with potentials will become useful in this context. \cite{Westermayr2021, Ceriotti2022}

Spectroscopic techniques more generally are widely used to characterize materials. Emerging ML models are being built that, for example, enable comparison with X-ray photoelectron spectroscopy data \cite{Golze2022} -- and even in the absence of those models, DFT-based predictions are often readily available (as long as a relatively small simulation cell is deemed sufficient). The application of such computational spectroscopy techniques to {\em validating} potentials against experiment is beginning to be explored. Shapeev and colleagues have shown how to validate ML potentials, in this case built using the MTP framework, against experimental extended X-ray absorption fine structure (EXAFS) data which provide a fingerprint of local structure. \cite{Shapeev2022} The authors found that a major source of possible discrepancies was the nature of the DFT reference data, which corresponds to the ``Quality of data labels'' point in Fig.\ \ref{fig:MLP_overview}b. 

For liquid and amorphous phases, structural information is difficult to obtain, and only typically accessible through indirect observations. A primary type of analysis involves inspecting the structure factor from diffraction and its Fourier transform which yields the pair distribution function (PDF). One key example is in the study of battery materials, such as nanoporous carbons, which can be experimentally characterized by PDF analysis (among other techniques). \cite{Forse2015} A recent ML-driven study compared computational predictions against those quantities. \cite{Wang2022a} We emphasize that these validations are typically about validating the structural model itself, not nuances of the potential. Nevertheless, the potential itself is influential in generating the structural model, so that the quality of the latter can act as a metric for the performance of a potential. We also emphasize that for amorphous materials, it will be particularly challenging to create accurate and reproducible {\em experimental} benchmarks, and it would be interesting to see whether more benchmarks of this type can be created in the future. \cite{Mata2017, Deringer2020b}

The computational speed of ML potentials makes it possible to generate structural models on the length scale of several nanometers and more, and with nanoscale structure. For example, such structural models can include grain boundaries and inhomogeneity arising from phase coexistence.
The accessibility of accurate large-scale simulations allows for convergence of structural metrics, such as the predicted structure factor, with system size. It is important to test this convergence, which is often not feasible with DFT, to ensure that any conclusions on the quality of the potential are reliable.

\section{Best-practice recommendations}

We suggest that the following should be included in publications that {\em introduce} an ML potential:

\begin{itemize}
    \item Energy and force errors from internal ($k$-fold) cross-validation, including a definition of how these errors have been obtained (RMSE or MAE, absolute force or force component error, etc.);
    \item Energy and force errors for one or more external (out-of-sample) test sets, for example, from separate MD simulations or random search;
    \item Comparison with experimental data wherever these are available from previous literature (Table \ref{tab:observables}), including a brief discussion of the errors or uncertainty of the literature values.
\end{itemize}

We suggest that the following should be included in publications that {\em use} an existing ML potential:

\begin{itemize}
    \item A mention of the above error metrics, if defined in the original publication;
    \item A brief comment on how, and to what extent, these previously given errors are applicable to the problem studied in the new work;
    \item If possible, benchmarks using a small-scale DFT simulation that is representative of the problem at hand (see Fig.\ \ref{fig:Si_JG} for one example);
    \item Wherever available, comparison with experimental data sourced from previous literature (Table \ref{tab:observables}).
\end{itemize}

We hope that these points will not only increase the confidence of the computational practitioner themselves (knowing that they are operating slightly away from the security of quantum mechanics), but also of experimental colleagues who will read the work. 

We also emphasize the importance, and the expected long-term advantages, of openly sharing training and benchmark data, as well as testing workflows especially as they become more complex. We refer again to recently published set of guidelines for ML models more generally, in Refs.\ \citenum{Artrith2021} and \citenum{Bender2022}, beyond the case of interatomic potentials discussed herein.

\section{Conclusions and outlook}

We have provided a tutorial-style overview of the multi-faceted problem of validating ML potential models for materials simulations. There has always been a need to ensure the validity and accuracy of interatomic potentials, and this need is becoming more acute as ML potentials are becoming increasingly widely used outside their specialized community of developers.

Looking forward, it would seem desirable to create openly available ``packaged'' tests that can be run directly from an openly available and easily accessible code, say, a Python script or automated workflow. The testing framework for elemental silicon described in Ref.\ \citenum{Bartok2018} is an excellent example of this, and we have benefited from it ourselves during the work described in Ref.\ \citenum{Morrow2022}. 

We expect that, with properly chosen reference configurations, numerical errors will remain important and indeed become more important. The question on which data exactly to carry out numerical validation, and whether there can be an optimized set of out-of-sample testing data for one material (or for many materials) constitutes a key challenge. 
We envision the increased use of random (RSS) configurations in this, as well as the creation and sharing of dedicated benchmark simulations, such as that in Fig.\ \ref{fig:Si_JG}. Once suitable structural snapshots are found which are representative of a given physical problem, anyone can download these structures, re-label them with the specific reference method used in their new potential, and evaluate the error on those. (The re-labeling step will be required in most cases, because usually ML potentials are fitted to data at different computational levels.) We believe that errors on a specified, physically-guided benchmark will be useful in evaluating future generations of ML potentials, and that they will convey information that simple cross-validation can not.

We hope that the ideas and approaches discussed in this tutorial will help to establish ML potentials as everyday tools in materials modeling, in the same way that DFT-based simulation methods are abundantly and very successfully used today. We look forward to seeing how, in the years ahead, carefully crafted and validated ML potentials will accelerate scientific discovery in physics, chemistry, and related fields.

\section*{Acknowledgments}

We thank J. George and Y. Zhou for helpful comments on the manuscript.
J.D.M. acknowledges funding from the EPSRC Centre for Doctoral Training in Inorganic Chemistry for Future Manufacturing (OxICFM), EP/S023828/1.
J.L.A.G. acknowledges a UKRI Linacre - The EPA Cephalosporin Scholarship, support from an EPSRC DTP award (EP/T517811/1), and from the Department of Chemistry, University of Oxford.
V.L.D. acknowledges a UK Research and Innovation Frontier Research grant [grant number EP/X016188/1].
Structural drawings were created with the help of OVITO. \cite{Stukowski2009}

\section*{Author Declarations}

\subsection*{Conflict of Interest}

The authors have no conflicts to disclose.

\section*{Data availability}

Data generated in this work, as well as Python code to reproduce relevant figures, will be provided openly upon journal publication.


\section*{References}

\begin{thebibliography}{81}%
\makeatletter
\providecommand \@ifxundefined [1]{%
 \@ifx{#1\undefined}
}%
\providecommand \@ifnum [1]{%
 \ifnum #1\expandafter \@firstoftwo
 \else \expandafter \@secondoftwo
 \fi
}%
\providecommand \@ifx [1]{%
 \ifx #1\expandafter \@firstoftwo
 \else \expandafter \@secondoftwo
 \fi
}%
\providecommand \natexlab [1]{#1}%
\providecommand \enquote  [1]{``#1''}%
\providecommand \bibnamefont  [1]{#1}%
\providecommand \bibfnamefont [1]{#1}%
\providecommand \citenamefont [1]{#1}%
\providecommand \href@noop [0]{\@secondoftwo}%
\providecommand \href [0]{\begingroup \@sanitize@url \@href}%
\providecommand \@href[1]{\@@startlink{#1}\@@href}%
\providecommand \@@href[1]{\endgroup#1\@@endlink}%
\providecommand \@sanitize@url [0]{\catcode `\\12\catcode `\$12\catcode
  `\&12\catcode `\#12\catcode `\^12\catcode `\_12\catcode `\%12\relax}%
\providecommand \@@startlink[1]{}%
\providecommand \@@endlink[0]{}%
\providecommand \url  [0]{\begingroup\@sanitize@url \@url }%
\providecommand \@url [1]{\endgroup\@href {#1}{\urlprefix }}%
\providecommand \urlprefix  [0]{URL }%
\providecommand \Eprint [0]{\href }%
\providecommand \doibase [0]{http://dx.doi.org/}%
\providecommand \selectlanguage [0]{\@gobble}%
\providecommand \bibinfo  [0]{\@secondoftwo}%
\providecommand \bibfield  [0]{\@secondoftwo}%
\providecommand \translation [1]{[#1]}%
\providecommand \BibitemOpen [0]{}%
\providecommand \bibitemStop [0]{}%
\providecommand \bibitemNoStop [0]{.\EOS\space}%
\providecommand \EOS [0]{\spacefactor3000\relax}%
\providecommand \BibitemShut  [1]{\csname bibitem#1\endcsname}%
\let\auto@bib@innerbib\@empty
%</preamble>
\bibitem [{\citenamefont {Behler}(2017)}]{Behler2017}%
  \BibitemOpen
  \bibfield  {author} {\bibinfo {author} {\bibfnamefont {J.}~\bibnamefont
  {Behler}},\ }\bibfield  {title} {\enquote {\bibinfo {title} {First principles
  neural network potentials for reactive simulations of large molecular and
  condensed systems},}\ }\href {\doibase 10.1002/anie.201703114} {\bibfield
  {journal} {\bibinfo  {journal} {Angew. Chem. Int. Ed.}\ }\textbf {\bibinfo
  {volume} {56}},\ \bibinfo {pages} {12828--12840} (\bibinfo {year}
  {2017})}\BibitemShut {NoStop}%
\bibitem [{\citenamefont {Deringer}, \citenamefont {Caro},\ and\ \citenamefont
  {Cs\'a{}nyi}(2019)}]{Deringer2019}%
  \BibitemOpen
  \bibfield  {author} {\bibinfo {author} {\bibfnamefont {V.~L.}\ \bibnamefont
  {Deringer}}, \bibinfo {author} {\bibfnamefont {M.~A.}\ \bibnamefont {Caro}},
  \ and\ \bibinfo {author} {\bibfnamefont {G.}~\bibnamefont {Cs\'a{}nyi}},\
  }\bibfield  {title} {\enquote {\bibinfo {title} {Machine learning interatomic
  potentials as emerging tools for materials science},}\ }\href {\doibase
  10.1002/adma.201902765} {\bibfield  {journal} {\bibinfo  {journal} {Adv.
  Mater.}\ }\textbf {\bibinfo {volume} {31}},\ \bibinfo {pages} {1902765}
  (\bibinfo {year} {2019})}\BibitemShut {NoStop}%
\bibitem [{\citenamefont {No\'e{}}\ \emph {et~al.}(2020)\citenamefont
  {No\'e{}}, \citenamefont {Tkatchenko}, \citenamefont {M\"u{}ller},\ and\
  \citenamefont {Clementi}}]{Noe2020}%
  \BibitemOpen
  \bibfield  {author} {\bibinfo {author} {\bibfnamefont {F.}~\bibnamefont
  {No\'e{}}}, \bibinfo {author} {\bibfnamefont {A.}~\bibnamefont {Tkatchenko}},
  \bibinfo {author} {\bibfnamefont {K.-R.}\ \bibnamefont {M\"u{}ller}}, \ and\
  \bibinfo {author} {\bibfnamefont {C.}~\bibnamefont {Clementi}},\ }\bibfield
  {title} {\enquote {\bibinfo {title} {Machine learning for molecular
  simulation},}\ }\href {\doibase 10.1146/annurev-physchem-042018-052331}
  {\bibfield  {journal} {\bibinfo  {journal} {Annu. Rev. Phys. Chem.}\ }\textbf
  {\bibinfo {volume} {71}},\ \bibinfo {pages} {361--390} (\bibinfo {year}
  {2020})}\BibitemShut {NoStop}%
\bibitem [{\citenamefont {Unke}\ \emph {et~al.}(2021)\citenamefont {Unke},
  \citenamefont {Chmiela}, \citenamefont {Sauceda}, \citenamefont {Gastegger},
  \citenamefont {Poltavsky}, \citenamefont {Sch\"u{}tt}, \citenamefont
  {Tkatchenko},\ and\ \citenamefont {M\"u{}ller}}]{Unke2021}%
  \BibitemOpen
  \bibfield  {author} {\bibinfo {author} {\bibfnamefont {O.~T.}\ \bibnamefont
  {Unke}}, \bibinfo {author} {\bibfnamefont {S.}~\bibnamefont {Chmiela}},
  \bibinfo {author} {\bibfnamefont {H.~E.}\ \bibnamefont {Sauceda}}, \bibinfo
  {author} {\bibfnamefont {M.}~\bibnamefont {Gastegger}}, \bibinfo {author}
  {\bibfnamefont {I.}~\bibnamefont {Poltavsky}}, \bibinfo {author}
  {\bibfnamefont {K.~T.}\ \bibnamefont {Sch\"u{}tt}}, \bibinfo {author}
  {\bibfnamefont {A.}~\bibnamefont {Tkatchenko}}, \ and\ \bibinfo {author}
  {\bibfnamefont {K.-R.}\ \bibnamefont {M\"u{}ller}},\ }\bibfield  {title}
  {\enquote {\bibinfo {title} {Machine learning force fields},}\ }\href
  {\doibase 10.1021/acs.chemrev.0c01111} {\bibfield  {journal} {\bibinfo
  {journal} {Chem. Rev.}\ }\textbf {\bibinfo {volume} {121}},\ \bibinfo {pages}
  {10142--10186} (\bibinfo {year} {2021})}\BibitemShut {NoStop}%
\bibitem [{\citenamefont {Friederich}\ \emph {et~al.}(2021)\citenamefont
  {Friederich}, \citenamefont {H\"a{}se}, \citenamefont {Proppe},\ and\
  \citenamefont {Aspuru-Guzik}}]{Friederich2021}%
  \BibitemOpen
  \bibfield  {author} {\bibinfo {author} {\bibfnamefont {P.}~\bibnamefont
  {Friederich}}, \bibinfo {author} {\bibfnamefont {F.}~\bibnamefont
  {H\"a{}se}}, \bibinfo {author} {\bibfnamefont {J.}~\bibnamefont {Proppe}}, \
  and\ \bibinfo {author} {\bibfnamefont {A.}~\bibnamefont {Aspuru-Guzik}},\
  }\bibfield  {title} {\enquote {\bibinfo {title} {Machine-learned potentials
  for next-generation matter simulations},}\ }\href {\doibase
  10.1038/s41563-020-0777-6} {\bibfield  {journal} {\bibinfo  {journal} {Nat.
  Mater.}\ }\textbf {\bibinfo {volume} {20}},\ \bibinfo {pages} {750--761}
  (\bibinfo {year} {2021})}\BibitemShut {NoStop}%
\bibitem [{\citenamefont {Cheng}\ \emph {et~al.}(2020)\citenamefont {Cheng},
  \citenamefont {Mazzola}, \citenamefont {Pickard},\ and\ \citenamefont
  {Ceriotti}}]{Cheng2020}%
  \BibitemOpen
  \bibfield  {author} {\bibinfo {author} {\bibfnamefont {B.}~\bibnamefont
  {Cheng}}, \bibinfo {author} {\bibfnamefont {G.}~\bibnamefont {Mazzola}},
  \bibinfo {author} {\bibfnamefont {C.~J.}\ \bibnamefont {Pickard}}, \ and\
  \bibinfo {author} {\bibfnamefont {M.}~\bibnamefont {Ceriotti}},\ }\bibfield
  {title} {\enquote {\bibinfo {title} {Evidence for supercritical behaviour of
  high-pressure liquid hydrogen},}\ }\href {\doibase 10.1038/s41586-020-2677-y}
  {\bibfield  {journal} {\bibinfo  {journal} {Nature}\ }\textbf {\bibinfo
  {volume} {585}},\ \bibinfo {pages} {217--220} (\bibinfo {year}
  {2020})}\BibitemShut {NoStop}%
\bibitem [{\citenamefont {Deringer}\ \emph
  {et~al.}(2021{\natexlab{a}})\citenamefont {Deringer}, \citenamefont
  {Bernstein}, \citenamefont {Cs\'anyi}, \citenamefont {Ben~Mahmoud},
  \citenamefont {Ceriotti}, \citenamefont {Wilson}, \citenamefont {Drabold},\
  and\ \citenamefont {Elliott}}]{Deringer2021}%
  \BibitemOpen
  \bibfield  {author} {\bibinfo {author} {\bibfnamefont {V.~L.}\ \bibnamefont
  {Deringer}}, \bibinfo {author} {\bibfnamefont {N.}~\bibnamefont {Bernstein}},
  \bibinfo {author} {\bibfnamefont {G.}~\bibnamefont {Cs\'anyi}}, \bibinfo
  {author} {\bibfnamefont {C.}~\bibnamefont {Ben~Mahmoud}}, \bibinfo {author}
  {\bibfnamefont {M.}~\bibnamefont {Ceriotti}}, \bibinfo {author}
  {\bibfnamefont {M.}~\bibnamefont {Wilson}}, \bibinfo {author} {\bibfnamefont
  {D.~A.}\ \bibnamefont {Drabold}}, \ and\ \bibinfo {author} {\bibfnamefont
  {S.~R.}\ \bibnamefont {Elliott}},\ }\bibfield  {title} {\enquote {\bibinfo
  {title} {Origins of structural and electronic transitions in disordered
  silicon},}\ }\href {\doibase 10.1038/s41586-020-03072-z} {\bibfield
  {journal} {\bibinfo  {journal} {Nature}\ }\textbf {\bibinfo {volume} {589}},\
  \bibinfo {pages} {59--64} (\bibinfo {year} {2021}{\natexlab{a}})}\BibitemShut
  {NoStop}%
\bibitem [{\citenamefont {Zong}\ \emph {et~al.}(2021)\citenamefont {Zong},
  \citenamefont {Robinson}, \citenamefont {A}, \citenamefont {Zhao},
  \citenamefont {Scandolo}, \citenamefont {Ding},\ and\ \citenamefont
  {Ackland}}]{Zong2021}%
  \BibitemOpen
  \bibfield  {author} {\bibinfo {author} {\bibfnamefont {H.}~\bibnamefont
  {Zong}}, \bibinfo {author} {\bibfnamefont {V.~N.}\ \bibnamefont {Robinson}},
  \bibinfo {author} {\bibfnamefont {H.}~\bibnamefont {A}}, \bibinfo {author}
  {\bibfnamefont {L.}~\bibnamefont {Zhao}}, \bibinfo {author} {\bibfnamefont
  {S.}~\bibnamefont {Scandolo}}, \bibinfo {author} {\bibfnamefont
  {X.}~\bibnamefont {Ding}}, \ and\ \bibinfo {author} {\bibfnamefont {G.~J.}\
  \bibnamefont {Ackland}},\ }\bibfield  {title} {\enquote {\bibinfo {title}
  {Free electron to electride transition in dense liquid potassium},}\ }\href
  {\doibase 10.1038/s41567-021-01244-w} {\bibfield  {journal} {\bibinfo
  {journal} {Nat. Phys.}\ }\textbf {\bibinfo {volume} {17}},\ \bibinfo {pages}
  {955--960} (\bibinfo {year} {2021})}\BibitemShut {NoStop}%
\bibitem [{\citenamefont {Caro}\ \emph {et~al.}(2018)\citenamefont {Caro},
  \citenamefont {Deringer}, \citenamefont {Koskinen}, \citenamefont {Laurila},\
  and\ \citenamefont {Cs\'anyi}}]{Caro2018}%
  \BibitemOpen
  \bibfield  {author} {\bibinfo {author} {\bibfnamefont {M.~A.}\ \bibnamefont
  {Caro}}, \bibinfo {author} {\bibfnamefont {V.~L.}\ \bibnamefont {Deringer}},
  \bibinfo {author} {\bibfnamefont {J.}~\bibnamefont {Koskinen}}, \bibinfo
  {author} {\bibfnamefont {T.}~\bibnamefont {Laurila}}, \ and\ \bibinfo
  {author} {\bibfnamefont {G.}~\bibnamefont {Cs\'anyi}},\ }\bibfield  {title}
  {\enquote {\bibinfo {title} {Growth mechanism and origin of high $s{p}^{3}$
  content in tetrahedral amorphous carbon},}\ }\href {\doibase
  10.1103/PhysRevLett.120.166101} {\bibfield  {journal} {\bibinfo  {journal}
  {Phys. Rev. Lett.}\ }\textbf {\bibinfo {volume} {120}},\ \bibinfo {pages}
  {166101} (\bibinfo {year} {2018})}\BibitemShut {NoStop}%
\bibitem [{\citenamefont {Westermayr}\ \emph {et~al.}(2022)\citenamefont
  {Westermayr}, \citenamefont {Gastegger}, \citenamefont {V\"o{}r\"o{}s},
  \citenamefont {Panzenboeck}, \citenamefont {Joerg}, \citenamefont
  {Gonz\'a{}lez},\ and\ \citenamefont {Marquetand}}]{Westermayr2022}%
  \BibitemOpen
  \bibfield  {author} {\bibinfo {author} {\bibfnamefont {J.}~\bibnamefont
  {Westermayr}}, \bibinfo {author} {\bibfnamefont {M.}~\bibnamefont
  {Gastegger}}, \bibinfo {author} {\bibfnamefont {D.}~\bibnamefont
  {V\"o{}r\"o{}s}}, \bibinfo {author} {\bibfnamefont {L.}~\bibnamefont
  {Panzenboeck}}, \bibinfo {author} {\bibfnamefont {F.}~\bibnamefont {Joerg}},
  \bibinfo {author} {\bibfnamefont {L.}~\bibnamefont {Gonz\'a{}lez}}, \ and\
  \bibinfo {author} {\bibfnamefont {P.}~\bibnamefont {Marquetand}},\ }\bibfield
   {title} {\enquote {\bibinfo {title} {Deep learning study of tyrosine reveals
  that roaming can lead to photodamage},}\ }\href {\doibase
  10.1038/s41557-022-00950-z} {\bibfield  {journal} {\bibinfo  {journal} {Nat.
  Chem.}\ }\textbf {\bibinfo {volume} {14}},\ \bibinfo {pages} {914--919}
  (\bibinfo {year} {2022})}\BibitemShut {NoStop}%
\bibitem [{\citenamefont {Sosso}\ \emph {et~al.}(2013)\citenamefont {Sosso},
  \citenamefont {Miceli}, \citenamefont {Caravati}, \citenamefont {Giberti},
  \citenamefont {Behler},\ and\ \citenamefont {Bernasconi}}]{Sosso2013}%
  \BibitemOpen
  \bibfield  {author} {\bibinfo {author} {\bibfnamefont {G.~C.}\ \bibnamefont
  {Sosso}}, \bibinfo {author} {\bibfnamefont {G.}~\bibnamefont {Miceli}},
  \bibinfo {author} {\bibfnamefont {S.}~\bibnamefont {Caravati}}, \bibinfo
  {author} {\bibfnamefont {F.}~\bibnamefont {Giberti}}, \bibinfo {author}
  {\bibfnamefont {J.}~\bibnamefont {Behler}}, \ and\ \bibinfo {author}
  {\bibfnamefont {M.}~\bibnamefont {Bernasconi}},\ }\bibfield  {title}
  {\enquote {\bibinfo {title} {Fast crystallization of the phase change
  compound {GeTe} by large-scale molecular dynamics simulations},}\ }\href
  {\doibase 10.1021/jz402268v} {\bibfield  {journal} {\bibinfo  {journal} {J.
  Phys. Chem. Lett.}\ }\textbf {\bibinfo {volume} {4}},\ \bibinfo {pages}
  {4241--4246} (\bibinfo {year} {2013})}\BibitemShut {NoStop}%
\bibitem [{\citenamefont {Konstantinou}\ \emph {et~al.}(2019)\citenamefont
  {Konstantinou}, \citenamefont {Mocanu}, \citenamefont {Lee},\ and\
  \citenamefont {Elliott}}]{Konstantinou2019}%
  \BibitemOpen
  \bibfield  {author} {\bibinfo {author} {\bibfnamefont {K.}~\bibnamefont
  {Konstantinou}}, \bibinfo {author} {\bibfnamefont {F.~C.}\ \bibnamefont
  {Mocanu}}, \bibinfo {author} {\bibfnamefont {T.-H.}\ \bibnamefont {Lee}}, \
  and\ \bibinfo {author} {\bibfnamefont {S.~R.}\ \bibnamefont {Elliott}},\
  }\bibfield  {title} {\enquote {\bibinfo {title} {Revealing the intrinsic
  nature of the mid-gap defects in amorphous {Ge$_{2}$Sb$_{2}$Te$_{5}$}},}\
  }\href {\doibase 10.1038/s41467-019-10980-w} {\bibfield  {journal} {\bibinfo
  {journal} {Nat. Commun.}\ }\textbf {\bibinfo {volume} {10}},\ \bibinfo
  {pages} {3065} (\bibinfo {year} {2019})}\BibitemShut {NoStop}%
\bibitem [{\citenamefont {Artrith}, \citenamefont {Urban},\ and\ \citenamefont
  {Ceder}(2018)}]{Artrith2018}%
  \BibitemOpen
  \bibfield  {author} {\bibinfo {author} {\bibfnamefont {N.}~\bibnamefont
  {Artrith}}, \bibinfo {author} {\bibfnamefont {A.}~\bibnamefont {Urban}}, \
  and\ \bibinfo {author} {\bibfnamefont {G.}~\bibnamefont {Ceder}},\ }\bibfield
   {title} {\enquote {\bibinfo {title} {Constructing first-principles phase
  diagrams of amorphous {Li}$_{x}${Si} using machine-learning-assisted sampling
  with an evolutionary algorithm},}\ }\href {\doibase 10.1063/1.5017661}
  {\bibfield  {journal} {\bibinfo  {journal} {J. Chem. Phys.}\ }\textbf
  {\bibinfo {volume} {148}},\ \bibinfo {pages} {241711} (\bibinfo {year}
  {2018})}\BibitemShut {NoStop}%
\bibitem [{\citenamefont {Wang}\ \emph {et~al.}(2020)\citenamefont {Wang},
  \citenamefont {Aoyagi}, \citenamefont {Wisesa},\ and\ \citenamefont
  {Mueller}}]{Wang2020b}%
  \BibitemOpen
  \bibfield  {author} {\bibinfo {author} {\bibfnamefont {C.}~\bibnamefont
  {Wang}}, \bibinfo {author} {\bibfnamefont {K.}~\bibnamefont {Aoyagi}},
  \bibinfo {author} {\bibfnamefont {P.}~\bibnamefont {Wisesa}}, \ and\ \bibinfo
  {author} {\bibfnamefont {T.}~\bibnamefont {Mueller}},\ }\bibfield  {title}
  {\enquote {\bibinfo {title} {{Lithium Ion Conduction in Cathode Coating
  Materials from On-the-Fly Machine Learning}},}\ }\href {\doibase
  10.1021/acs.chemmater.9b04663} {\bibfield  {journal} {\bibinfo  {journal}
  {Chem. Mater.}\ }\textbf {\bibinfo {volume} {32}},\ \bibinfo {pages}
  {3741--3752} (\bibinfo {year} {2020})}\BibitemShut {NoStop}%
\bibitem [{\citenamefont {Wang}\ \emph
  {et~al.}(2022{\natexlab{a}})\citenamefont {Wang}, \citenamefont {Panchal},
  \citenamefont {Gautam},\ and\ \citenamefont {Canepa}}]{Wang2022}%
  \BibitemOpen
  \bibfield  {author} {\bibinfo {author} {\bibfnamefont {J.}~\bibnamefont
  {Wang}}, \bibinfo {author} {\bibfnamefont {A.~A.}\ \bibnamefont {Panchal}},
  \bibinfo {author} {\bibfnamefont {G.~S.}\ \bibnamefont {Gautam}}, \ and\
  \bibinfo {author} {\bibfnamefont {P.}~\bibnamefont {Canepa}},\ }\bibfield
  {title} {\enquote {\bibinfo {title} {The resistive nature of decomposing
  interfaces of solid electrolytes with alkali metal electrodes},}\ }\href
  {\doibase 10.1039/D2TA02202H} {\bibfield  {journal} {\bibinfo  {journal} {J.
  Mater. Chem. A}\ }\textbf {\bibinfo {volume} {10}},\ \bibinfo {pages}
  {19732--19742} (\bibinfo {year} {2022}{\natexlab{a}})}\BibitemShut {NoStop}%
\bibitem [{\citenamefont {Staacke}\ \emph {et~al.}(2022)\citenamefont
  {Staacke}, \citenamefont {Huss}, \citenamefont {Margraf}, \citenamefont
  {Reuter},\ and\ \citenamefont {Scheurer}}]{Staacke2022}%
  \BibitemOpen
  \bibfield  {author} {\bibinfo {author} {\bibfnamefont {C.~G.}\ \bibnamefont
  {Staacke}}, \bibinfo {author} {\bibfnamefont {T.}~\bibnamefont {Huss}},
  \bibinfo {author} {\bibfnamefont {J.~T.}\ \bibnamefont {Margraf}}, \bibinfo
  {author} {\bibfnamefont {K.}~\bibnamefont {Reuter}}, \ and\ \bibinfo {author}
  {\bibfnamefont {C.}~\bibnamefont {Scheurer}},\ }\bibfield  {title} {\enquote
  {\bibinfo {title} {Tackling structural complexity in
  {Li}$_{2}${S}-{P}$_{2}${S}$_{5}$ solid-state electrolytes using machine
  learning potentials},}\ }\href {\doibase 10.3390/nano12172950} {\bibfield
  {journal} {\bibinfo  {journal} {Nanomaterials}\ }\textbf {\bibinfo {volume}
  {12}},\ \bibinfo {pages} {2950} (\bibinfo {year} {2022})}\BibitemShut
  {NoStop}%
\bibitem [{\citenamefont {Gubaev}\ \emph {et~al.}(2019)\citenamefont {Gubaev},
  \citenamefont {Podryabinkin}, \citenamefont {Hart},\ and\ \citenamefont
  {Shapeev}}]{Gubaev2019}%
  \BibitemOpen
  \bibfield  {author} {\bibinfo {author} {\bibfnamefont {K.}~\bibnamefont
  {Gubaev}}, \bibinfo {author} {\bibfnamefont {E.~V.}\ \bibnamefont
  {Podryabinkin}}, \bibinfo {author} {\bibfnamefont {G.~L.~W.}\ \bibnamefont
  {Hart}}, \ and\ \bibinfo {author} {\bibfnamefont {A.~V.}\ \bibnamefont
  {Shapeev}},\ }\bibfield  {title} {\enquote {\bibinfo {title} {{Accelerating
  High-Throughput Searches for New Alloys with Active Learning of Interatomic
  Potentials}},}\ }\href@noop {} {\bibfield  {journal} {\bibinfo  {journal}
  {Comput. Mater. Sci.}\ }\textbf {\bibinfo {volume} {156}},\ \bibinfo {pages}
  {148--156} (\bibinfo {year} {2019})}\BibitemShut {NoStop}%
\bibitem [{\citenamefont {Marchand}\ and\ \citenamefont
  {Curtin}(2022)}]{Marchand2022}%
  \BibitemOpen
  \bibfield  {author} {\bibinfo {author} {\bibfnamefont {D.}~\bibnamefont
  {Marchand}}\ and\ \bibinfo {author} {\bibfnamefont {W.~A.}\ \bibnamefont
  {Curtin}},\ }\bibfield  {title} {\enquote {\bibinfo {title} {Machine learning
  for metallurgy {IV}: A neural network potential for {Al}-{Cu}-{Mg} and
  {Al}-{Cu}-{Mg}-{Zn}},}\ }\href {\doibase 10.1103/PhysRevMaterials.6.053803}
  {\bibfield  {journal} {\bibinfo  {journal} {Phys. Rev. Mater.}\ }\textbf
  {\bibinfo {volume} {6}},\ \bibinfo {pages} {053803} (\bibinfo {year}
  {2022})}\BibitemShut {NoStop}%
\bibitem [{\citenamefont {Zuo}\ \emph {et~al.}(2020)\citenamefont {Zuo},
  \citenamefont {Chen}, \citenamefont {Li}, \citenamefont {Deng}, \citenamefont
  {Chen}, \citenamefont {Behler}, \citenamefont {Cs\'a{}nyi}, \citenamefont
  {Shapeev}, \citenamefont {Thompson}, \citenamefont {Wood},\ and\
  \citenamefont {Ong}}]{Zuo2020}%
  \BibitemOpen
  \bibfield  {author} {\bibinfo {author} {\bibfnamefont {Y.}~\bibnamefont
  {Zuo}}, \bibinfo {author} {\bibfnamefont {C.}~\bibnamefont {Chen}}, \bibinfo
  {author} {\bibfnamefont {X.}~\bibnamefont {Li}}, \bibinfo {author}
  {\bibfnamefont {Z.}~\bibnamefont {Deng}}, \bibinfo {author} {\bibfnamefont
  {Y.}~\bibnamefont {Chen}}, \bibinfo {author} {\bibfnamefont {J.}~\bibnamefont
  {Behler}}, \bibinfo {author} {\bibfnamefont {G.}~\bibnamefont {Cs\'a{}nyi}},
  \bibinfo {author} {\bibfnamefont {A.~V.}\ \bibnamefont {Shapeev}}, \bibinfo
  {author} {\bibfnamefont {A.~P.}\ \bibnamefont {Thompson}}, \bibinfo {author}
  {\bibfnamefont {M.~A.}\ \bibnamefont {Wood}}, \ and\ \bibinfo {author}
  {\bibfnamefont {S.~P.}\ \bibnamefont {Ong}},\ }\bibfield  {title} {\enquote
  {\bibinfo {title} {Performance and cost assessment of machine learning
  interatomic potentials},}\ }\href {\doibase 10.1021/acs.jpca.9b08723}
  {\bibfield  {journal} {\bibinfo  {journal} {J. Phys. Chem. A}\ }\textbf
  {\bibinfo {volume} {124}},\ \bibinfo {pages} {731--745} (\bibinfo {year}
  {2020})}\BibitemShut {NoStop}%
\bibitem [{\citenamefont {Rowe}\ \emph {et~al.}(2020)\citenamefont {Rowe},
  \citenamefont {Deringer}, \citenamefont {Gasparotto}, \citenamefont
  {Cs\'anyi},\ and\ \citenamefont {Michaelides}}]{Rowe2020}%
  \BibitemOpen
  \bibfield  {author} {\bibinfo {author} {\bibfnamefont {P.}~\bibnamefont
  {Rowe}}, \bibinfo {author} {\bibfnamefont {V.~L.}\ \bibnamefont {Deringer}},
  \bibinfo {author} {\bibfnamefont {P.}~\bibnamefont {Gasparotto}}, \bibinfo
  {author} {\bibfnamefont {G.}~\bibnamefont {Cs\'anyi}}, \ and\ \bibinfo
  {author} {\bibfnamefont {A.}~\bibnamefont {Michaelides}},\ }\bibfield
  {title} {\enquote {\bibinfo {title} {An accurate and transferable machine
  learning potential for carbon},}\ }\href {\doibase 10.1063/5.0005084}
  {\bibfield  {journal} {\bibinfo  {journal} {J. Chem. Phys.}\ }\textbf
  {\bibinfo {volume} {153}},\ \bibinfo {pages} {034702} (\bibinfo {year}
  {2020})}\BibitemShut {NoStop}%
\bibitem [{\citenamefont {Qian}\ \emph {et~al.}(2021)\citenamefont {Qian},
  \citenamefont {McLean}, \citenamefont {Hedman},\ and\ \citenamefont
  {Ding}}]{Qian2021}%
  \BibitemOpen
  \bibfield  {author} {\bibinfo {author} {\bibfnamefont {C.}~\bibnamefont
  {Qian}}, \bibinfo {author} {\bibfnamefont {B.}~\bibnamefont {McLean}},
  \bibinfo {author} {\bibfnamefont {D.}~\bibnamefont {Hedman}}, \ and\ \bibinfo
  {author} {\bibfnamefont {F.}~\bibnamefont {Ding}},\ }\bibfield  {title}
  {\enquote {\bibinfo {title} {A comprehensive assessment of empirical
  potentials for carbon materials},}\ }\href {\doibase 10.1063/5.0052870}
  {\bibfield  {journal} {\bibinfo  {journal} {APL Mater.}\ }\textbf {\bibinfo
  {volume} {9}},\ \bibinfo {pages} {061102} (\bibinfo {year}
  {2021})}\BibitemShut {NoStop}%
\bibitem [{\citenamefont {Aghajamali}\ and\ \citenamefont
  {Karton}(2021)}]{Aghajamali2021}%
  \BibitemOpen
  \bibfield  {author} {\bibinfo {author} {\bibfnamefont {A.}~\bibnamefont
  {Aghajamali}}\ and\ \bibinfo {author} {\bibfnamefont {A.}~\bibnamefont
  {Karton}},\ }\bibfield  {title} {\enquote {\bibinfo {title} {Can force fields
  developed for carbon nanomaterials describe the isomerization energies of
  fullerenes?}}\ }\href {\doibase 10.1016/j.cplett.2021.138853} {\bibfield
  {journal} {\bibinfo  {journal} {Chem. Phys. Lett.}\ }\textbf {\bibinfo
  {volume} {779}},\ \bibinfo {pages} {138853} (\bibinfo {year}
  {2021})}\BibitemShut {NoStop}%
\bibitem [{\citenamefont {Qamar}\ \emph {et~al.}(2022)\citenamefont {Qamar},
  \citenamefont {Mrovec}, \citenamefont {Lysogorskiy}, \citenamefont
  {Bochkarev},\ and\ \citenamefont {Drautz}}]{Qamar2022}%
  \BibitemOpen
  \bibfield  {author} {\bibinfo {author} {\bibfnamefont {M.}~\bibnamefont
  {Qamar}}, \bibinfo {author} {\bibfnamefont {M.}~\bibnamefont {Mrovec}},
  \bibinfo {author} {\bibfnamefont {Y.}~\bibnamefont {Lysogorskiy}}, \bibinfo
  {author} {\bibfnamefont {A.}~\bibnamefont {Bochkarev}}, \ and\ \bibinfo
  {author} {\bibfnamefont {R.}~\bibnamefont {Drautz}},\ }\bibfield  {title}
  {\enquote {\bibinfo {title} {Atomic cluster expansion for quantum-accurate
  large-scale simulations of carbon},}\ }\href {\doibase
  https://doi.org/10.48550/arXiv.2210.09161} {\bibfield  {journal} {\bibinfo
  {journal} {arXiv preprint\!\!}\ ,\ \bibinfo {pages} {arXiv:2210.09161
  [cond--mat.mtrl--sci]}} (\bibinfo {year} {2022})}\BibitemShut {NoStop}%
\bibitem [{\citenamefont {{de Tomas}}, \citenamefont {Suarez-Martinez},\ and\
  \citenamefont {Marks}(2016)}]{DeTomas2016}%
  \BibitemOpen
  \bibfield  {author} {\bibinfo {author} {\bibfnamefont {C.}~\bibnamefont {{de
  Tomas}}}, \bibinfo {author} {\bibfnamefont {I.}~\bibnamefont
  {Suarez-Martinez}}, \ and\ \bibinfo {author} {\bibfnamefont {N.~A.}\
  \bibnamefont {Marks}},\ }\bibfield  {title} {\enquote {\bibinfo {title}
  {Graphitization of amorphous carbons: A comparative study of interatomic
  potentials},}\ }\href {\doibase 10.1016/j.carbon.2016.08.024} {\bibfield
  {journal} {\bibinfo  {journal} {Carbon}\ }\textbf {\bibinfo {volume} {109}},\
  \bibinfo {pages} {681--693} (\bibinfo {year} {2016})}\BibitemShut {NoStop}%
\bibitem [{\citenamefont {de~Tomas}\ \emph {et~al.}(2019)\citenamefont
  {de~Tomas}, \citenamefont {Aghajamali}, \citenamefont {Jones}, \citenamefont
  {Lim}, \citenamefont {L\'o{}pez}, \citenamefont {Suarez-Martinez},\ and\
  \citenamefont {Marks}}]{DeTomas2019}%
  \BibitemOpen
  \bibfield  {author} {\bibinfo {author} {\bibfnamefont {C.}~\bibnamefont
  {de~Tomas}}, \bibinfo {author} {\bibfnamefont {A.}~\bibnamefont
  {Aghajamali}}, \bibinfo {author} {\bibfnamefont {J.~L.}\ \bibnamefont
  {Jones}}, \bibinfo {author} {\bibfnamefont {D.~J.}\ \bibnamefont {Lim}},
  \bibinfo {author} {\bibfnamefont {M.~J.}\ \bibnamefont {L\'o{}pez}}, \bibinfo
  {author} {\bibfnamefont {I.}~\bibnamefont {Suarez-Martinez}}, \ and\ \bibinfo
  {author} {\bibfnamefont {N.~A.}\ \bibnamefont {Marks}},\ }\bibfield  {title}
  {\enquote {\bibinfo {title} {Transferability in interatomic potentials for
  carbon},}\ }\href {\doibase https://doi.org/10.1016/j.carbon.2019.07.074}
  {\bibfield  {journal} {\bibinfo  {journal} {Carbon}\ }\textbf {\bibinfo
  {volume} {155}},\ \bibinfo {pages} {624--634} (\bibinfo {year}
  {2019})}\BibitemShut {NoStop}%
\bibitem [{\citenamefont {George}\ \emph {et~al.}(2020)\citenamefont {George},
  \citenamefont {Hautier}, \citenamefont {Bart\'ok}, \citenamefont {Cs\'anyi},\
  and\ \citenamefont {Deringer}}]{George2020}%
  \BibitemOpen
  \bibfield  {author} {\bibinfo {author} {\bibfnamefont {J.}~\bibnamefont
  {George}}, \bibinfo {author} {\bibfnamefont {G.}~\bibnamefont {Hautier}},
  \bibinfo {author} {\bibfnamefont {A.~P.}\ \bibnamefont {Bart\'ok}}, \bibinfo
  {author} {\bibfnamefont {G.}~\bibnamefont {Cs\'anyi}}, \ and\ \bibinfo
  {author} {\bibfnamefont {V.~L.}\ \bibnamefont {Deringer}},\ }\bibfield
  {title} {\enquote {\bibinfo {title} {Combining phonon accuracy with high
  transferability in gaussian approximation potential models},}\ }\href
  {\doibase 10.1063/5.0013826} {\bibfield  {journal} {\bibinfo  {journal} {J.
  Chem. Phys.}\ }\textbf {\bibinfo {volume} {153}},\ \bibinfo {pages} {044104}
  (\bibinfo {year} {2020})}\BibitemShut {NoStop}%
\bibitem [{\citenamefont {Kov\'a{}cs}\ \emph {et~al.}(2021)\citenamefont
  {Kov\'a{}cs}, \citenamefont {{van der Oord}}, \citenamefont {Kucera},
  \citenamefont {Allen}, \citenamefont {Cole}, \citenamefont {Ortner},\ and\
  \citenamefont {Cs\'a{}nyi}}]{Kovacs2021}%
  \BibitemOpen
  \bibfield  {author} {\bibinfo {author} {\bibfnamefont {D.~P.}\ \bibnamefont
  {Kov\'a{}cs}}, \bibinfo {author} {\bibfnamefont {C.}~\bibnamefont {{van der
  Oord}}}, \bibinfo {author} {\bibfnamefont {J.}~\bibnamefont {Kucera}},
  \bibinfo {author} {\bibfnamefont {A.~E.~A.}\ \bibnamefont {Allen}}, \bibinfo
  {author} {\bibfnamefont {D.~J.}\ \bibnamefont {Cole}}, \bibinfo {author}
  {\bibfnamefont {C.}~\bibnamefont {Ortner}}, \ and\ \bibinfo {author}
  {\bibfnamefont {G.}~\bibnamefont {Cs\'a{}nyi}},\ }\bibfield  {title}
  {\enquote {\bibinfo {title} {Linear atomic cluster expansion force fields for
  organic molecules: Beyond {RMSE}},}\ }\href {\doibase
  10.1021/acs.jctc.1c00647} {\bibfield  {journal} {\bibinfo  {journal} {J.
  Chem. Theory Comput.}\ }\textbf {\bibinfo {volume} {17}},\ \bibinfo {pages}
  {7696--7711} (\bibinfo {year} {2021})}\BibitemShut {NoStop}%
\bibitem [{\citenamefont {Fu}\ \emph {et~al.}(2022)\citenamefont {Fu},
  \citenamefont {Wu}, \citenamefont {Wang}, \citenamefont {Xie}, \citenamefont
  {Keten}, \citenamefont {Gomez-Bombarelli},\ and\ \citenamefont
  {Jaakkola}}]{Fu2022}%
  \BibitemOpen
  \bibfield  {author} {\bibinfo {author} {\bibfnamefont {X.}~\bibnamefont
  {Fu}}, \bibinfo {author} {\bibfnamefont {Z.}~\bibnamefont {Wu}}, \bibinfo
  {author} {\bibfnamefont {W.}~\bibnamefont {Wang}}, \bibinfo {author}
  {\bibfnamefont {T.}~\bibnamefont {Xie}}, \bibinfo {author} {\bibfnamefont
  {S.}~\bibnamefont {Keten}}, \bibinfo {author} {\bibfnamefont
  {R.}~\bibnamefont {Gomez-Bombarelli}}, \ and\ \bibinfo {author}
  {\bibfnamefont {T.}~\bibnamefont {Jaakkola}},\ }\bibfield  {title} {\enquote
  {\bibinfo {title} {Forces are not enough: Benchmark and critical evaluation
  for machine learning force fields with molecular simulations},}\ }\href
  {\doibase https://doi.org/10.48550/arXiv.2210.07237} {\bibfield  {journal}
  {\bibinfo  {journal} {arXiv preprint\!\!}\ ,\ \bibinfo {pages}
  {arXiv:2210.07237 [physics.comp--ph]}} (\bibinfo {year} {2022})}\BibitemShut
  {NoStop}%
\bibitem [{\citenamefont {Vishwakarma}, \citenamefont {Sonpal},\ and\
  \citenamefont {Hachmann}(2021)}]{Vishvakarma2021}%
  \BibitemOpen
  \bibfield  {author} {\bibinfo {author} {\bibfnamefont {G.}~\bibnamefont
  {Vishwakarma}}, \bibinfo {author} {\bibfnamefont {A.}~\bibnamefont {Sonpal}},
  \ and\ \bibinfo {author} {\bibfnamefont {J.}~\bibnamefont {Hachmann}},\
  }\bibfield  {title} {\enquote {\bibinfo {title} {Metrics for benchmarking and
  uncertainty quantification: Quality, applicability, and best practices for
  machine learning in chemistry},}\ }\href {\doibase
  https://doi.org/10.1016/j.trechm.2020.12.004} {\bibfield  {journal} {\bibinfo
   {journal} {Trends Chem.}\ }\textbf {\bibinfo {volume} {3}},\ \bibinfo
  {pages} {146--156} (\bibinfo {year} {2021})}\BibitemShut {NoStop}%
\bibitem [{\citenamefont {Artrith}\ \emph {et~al.}(2021)\citenamefont
  {Artrith}, \citenamefont {Butler}, \citenamefont {Coudert}, \citenamefont
  {Han}, \citenamefont {Isayev}, \citenamefont {Jain},\ and\ \citenamefont
  {Walsh}}]{Artrith2021}%
  \BibitemOpen
  \bibfield  {author} {\bibinfo {author} {\bibfnamefont {N.}~\bibnamefont
  {Artrith}}, \bibinfo {author} {\bibfnamefont {K.~T.}\ \bibnamefont {Butler}},
  \bibinfo {author} {\bibfnamefont {F.-X.}\ \bibnamefont {Coudert}}, \bibinfo
  {author} {\bibfnamefont {S.}~\bibnamefont {Han}}, \bibinfo {author}
  {\bibfnamefont {O.}~\bibnamefont {Isayev}}, \bibinfo {author} {\bibfnamefont
  {A.}~\bibnamefont {Jain}}, \ and\ \bibinfo {author} {\bibfnamefont
  {A.}~\bibnamefont {Walsh}},\ }\bibfield  {title} {\enquote {\bibinfo {title}
  {Best practices in machine learning for chemistry},}\ }\href {\doibase
  10.1038/s41557-021-00716-z} {\bibfield  {journal} {\bibinfo  {journal} {Nat.
  Chem.}\ }\textbf {\bibinfo {volume} {13}},\ \bibinfo {pages} {505--508}
  (\bibinfo {year} {2021})}\BibitemShut {NoStop}%
\bibitem [{\citenamefont {Bender}\ \emph {et~al.}(2022)\citenamefont {Bender},
  \citenamefont {Schneider}, \citenamefont {Segler}, \citenamefont
  {Patrick~Walters}, \citenamefont {Engkvist},\ and\ \citenamefont
  {Rodrigues}}]{Bender2022}%
  \BibitemOpen
  \bibfield  {author} {\bibinfo {author} {\bibfnamefont {A.}~\bibnamefont
  {Bender}}, \bibinfo {author} {\bibfnamefont {N.}~\bibnamefont {Schneider}},
  \bibinfo {author} {\bibfnamefont {M.}~\bibnamefont {Segler}}, \bibinfo
  {author} {\bibfnamefont {W.}~\bibnamefont {Patrick~Walters}}, \bibinfo
  {author} {\bibfnamefont {O.}~\bibnamefont {Engkvist}}, \ and\ \bibinfo
  {author} {\bibfnamefont {T.}~\bibnamefont {Rodrigues}},\ }\bibfield  {title}
  {\enquote {\bibinfo {title} {Evaluation guidelines for machine learning tools
  in the chemical sciences},}\ }\href {\doibase 10.1038/s41570-022-00391-9}
  {\bibfield  {journal} {\bibinfo  {journal} {Nat. Rev. Chem.}\ }\textbf
  {\bibinfo {volume} {6}},\ \bibinfo {pages} {428--442} (\bibinfo {year}
  {2022})}\BibitemShut {NoStop}%
\bibitem [{\citenamefont {Miksch}\ \emph {et~al.}(2021)\citenamefont {Miksch},
  \citenamefont {Morawietz}, \citenamefont {K\"a{}stner}, \citenamefont
  {Urban},\ and\ \citenamefont {Artrith}}]{Miksch2021}%
  \BibitemOpen
  \bibfield  {author} {\bibinfo {author} {\bibfnamefont {A.~M.}\ \bibnamefont
  {Miksch}}, \bibinfo {author} {\bibfnamefont {T.}~\bibnamefont {Morawietz}},
  \bibinfo {author} {\bibfnamefont {J.}~\bibnamefont {K\"a{}stner}}, \bibinfo
  {author} {\bibfnamefont {A.}~\bibnamefont {Urban}}, \ and\ \bibinfo {author}
  {\bibfnamefont {N.}~\bibnamefont {Artrith}},\ }\bibfield  {title} {\enquote
  {\bibinfo {title} {Strategies for the construction of machine-learning
  potentials for accurate and efficient atomic-scale simulations},}\ }\href
  {\doibase 10.1088/2632-2153/abfd96} {\bibfield  {journal} {\bibinfo
  {journal} {Mach. Learn.: Sci. Technol.}\ }\textbf {\bibinfo {volume} {2}},\
  \bibinfo {pages} {031001} (\bibinfo {year} {2021})}\BibitemShut {NoStop}%
\bibitem [{\citenamefont {Deringer}(2020)}]{Deringer2020b}%
  \BibitemOpen
  \bibfield  {author} {\bibinfo {author} {\bibfnamefont {V.~L.}\ \bibnamefont
  {Deringer}},\ }\bibfield  {title} {\enquote {\bibinfo {title} {Modelling and
  understanding battery materials with machine-learning-driven atomistic
  simulations},}\ }\href {\doibase 10.1088/2515-7655/abb011} {\bibfield
  {journal} {\bibinfo  {journal} {J. Phys. Energy}\ }\textbf {\bibinfo {volume}
  {2}},\ \bibinfo {pages} {041003} (\bibinfo {year} {2020})}\BibitemShut
  {NoStop}%
\bibitem [{\citenamefont {Liu}\ \emph {et~al.}(2022)\citenamefont {Liu},
  \citenamefont {Verdi}, \citenamefont {Karsai},\ and\ \citenamefont
  {Kresse}}]{Liu2022}%
  \BibitemOpen
  \bibfield  {author} {\bibinfo {author} {\bibfnamefont {P.}~\bibnamefont
  {Liu}}, \bibinfo {author} {\bibfnamefont {C.}~\bibnamefont {Verdi}}, \bibinfo
  {author} {\bibfnamefont {F.}~\bibnamefont {Karsai}}, \ and\ \bibinfo {author}
  {\bibfnamefont {G.}~\bibnamefont {Kresse}},\ }\bibfield  {title} {\enquote
  {\bibinfo {title} {Phase transitions of zirconia: Machine-learned force
  fields beyond density functional theory},}\ }\href {\doibase
  10.1103/PhysRevB.105.L060102} {\bibfield  {journal} {\bibinfo  {journal}
  {Phys. Rev. B}\ }\textbf {\bibinfo {volume} {105}},\ \bibinfo {pages}
  {L060102} (\bibinfo {year} {2022})}\BibitemShut {NoStop}%
\bibitem [{\citenamefont {Podryabinkin}\ and\ \citenamefont
  {Shapeev}(2017)}]{Podryabinkin2017}%
  \BibitemOpen
  \bibfield  {author} {\bibinfo {author} {\bibfnamefont {E.~V.}\ \bibnamefont
  {Podryabinkin}}\ and\ \bibinfo {author} {\bibfnamefont {A.~V.}\ \bibnamefont
  {Shapeev}},\ }\bibfield  {title} {\enquote {\bibinfo {title} {Active learning
  of linearly parametrized interatomic potentials},}\ }\href {\doibase
  10.1016/j.commatsci.2017.08.031} {\bibfield  {journal} {\bibinfo  {journal}
  {Comput. Mater. Sci.}\ }\textbf {\bibinfo {volume} {140}},\ \bibinfo {pages}
  {171--180} (\bibinfo {year} {2017})}\BibitemShut {NoStop}%
\bibitem [{\citenamefont {Zhang}\ \emph {et~al.}(2019)\citenamefont {Zhang},
  \citenamefont {Lin}, \citenamefont {Wang}, \citenamefont {Car},\ and\
  \citenamefont {E}}]{Zhang2018}%
  \BibitemOpen
  \bibfield  {author} {\bibinfo {author} {\bibfnamefont {L.}~\bibnamefont
  {Zhang}}, \bibinfo {author} {\bibfnamefont {D.-Y.}\ \bibnamefont {Lin}},
  \bibinfo {author} {\bibfnamefont {H.}~\bibnamefont {Wang}}, \bibinfo {author}
  {\bibfnamefont {R.}~\bibnamefont {Car}}, \ and\ \bibinfo {author}
  {\bibfnamefont {W.}~\bibnamefont {E}},\ }\bibfield  {title} {\enquote
  {\bibinfo {title} {Active learning of uniformly accurate interatomic
  potentials for materials simulation},}\ }\href {\doibase
  10.1103/PhysRevMaterials.3.023804} {\bibfield  {journal} {\bibinfo  {journal}
  {Phys. Rev. Mater.}\ }\textbf {\bibinfo {volume} {3}},\ \bibinfo {pages}
  {023804} (\bibinfo {year} {2019})}\BibitemShut {NoStop}%
\bibitem [{\citenamefont {Vandermause}\ \emph {et~al.}(2020)\citenamefont
  {Vandermause}, \citenamefont {Torrisi}, \citenamefont {Batzner},
  \citenamefont {Xie}, \citenamefont {Sun}, \citenamefont {Kolpak},\ and\
  \citenamefont {Kozinsky}}]{Kozinsky2020}%
  \BibitemOpen
  \bibfield  {author} {\bibinfo {author} {\bibfnamefont {J.}~\bibnamefont
  {Vandermause}}, \bibinfo {author} {\bibfnamefont {S.~B.}\ \bibnamefont
  {Torrisi}}, \bibinfo {author} {\bibfnamefont {S.}~\bibnamefont {Batzner}},
  \bibinfo {author} {\bibfnamefont {Y.}~\bibnamefont {Xie}}, \bibinfo {author}
  {\bibfnamefont {L.}~\bibnamefont {Sun}}, \bibinfo {author} {\bibfnamefont
  {A.~M.}\ \bibnamefont {Kolpak}}, \ and\ \bibinfo {author} {\bibfnamefont
  {B.}~\bibnamefont {Kozinsky}},\ }\bibfield  {title} {\enquote {\bibinfo
  {title} {On-the-fly active learning of interpretable {Bayesian} force fields
  for atomistic rare events},}\ }\href {\doibase 10.1038/s41524-020-0283-z}
  {\bibfield  {journal} {\bibinfo  {journal} {npj Comput. Mater.}\ }\textbf
  {\bibinfo {volume} {6}},\ \bibinfo {pages} {20} (\bibinfo {year}
  {2020})}\BibitemShut {NoStop}%
\bibitem [{\citenamefont {Deringer}\ \emph
  {et~al.}(2021{\natexlab{b}})\citenamefont {Deringer}, \citenamefont
  {Bart\'o{}k}, \citenamefont {Bernstein}, \citenamefont {Wilkins},
  \citenamefont {Ceriotti},\ and\ \citenamefont {Cs\'a{}nyi}}]{Deringer2021a}%
  \BibitemOpen
  \bibfield  {author} {\bibinfo {author} {\bibfnamefont {V.~L.}\ \bibnamefont
  {Deringer}}, \bibinfo {author} {\bibfnamefont {A.~P.}\ \bibnamefont
  {Bart\'o{}k}}, \bibinfo {author} {\bibfnamefont {N.}~\bibnamefont
  {Bernstein}}, \bibinfo {author} {\bibfnamefont {D.~M.}\ \bibnamefont
  {Wilkins}}, \bibinfo {author} {\bibfnamefont {M.}~\bibnamefont {Ceriotti}}, \
  and\ \bibinfo {author} {\bibfnamefont {G.}~\bibnamefont {Cs\'a{}nyi}},\
  }\bibfield  {title} {\enquote {\bibinfo {title} {Gaussian process regression
  for materials and molecules},}\ }\href {\doibase 10.1021/acs.chemrev.1c00022}
  {\bibfield  {journal} {\bibinfo  {journal} {Chem. Rev.}\ }\textbf {\bibinfo
  {volume} {121}},\ \bibinfo {pages} {10073--10141} (\bibinfo {year}
  {2021}{\natexlab{b}})}\BibitemShut {NoStop}%
\bibitem [{\citenamefont {Bayerl}\ \emph {et~al.}(2022)\citenamefont {Bayerl},
  \citenamefont {Andolina}, \citenamefont {Dwaraknath},\ and\ \citenamefont
  {Saidi}}]{Bayerl2022}%
  \BibitemOpen
  \bibfield  {author} {\bibinfo {author} {\bibfnamefont {D.}~\bibnamefont
  {Bayerl}}, \bibinfo {author} {\bibfnamefont {C.~M.}\ \bibnamefont
  {Andolina}}, \bibinfo {author} {\bibfnamefont {S.}~\bibnamefont
  {Dwaraknath}}, \ and\ \bibinfo {author} {\bibfnamefont {W.~A.}\ \bibnamefont
  {Saidi}},\ }\bibfield  {title} {\enquote {\bibinfo {title} {Convergence
  acceleration in machine learning potentials for atomistic simulations},}\
  }\href {\doibase 10.1039/D1DD00005E} {\bibfield  {journal} {\bibinfo
  {journal} {Digital Discovery}\ }\textbf {\bibinfo {volume} {1}},\ \bibinfo
  {pages} {61--69} (\bibinfo {year} {2022})}\BibitemShut {NoStop}%
\bibitem [{\citenamefont {Behler}\ and\ \citenamefont
  {Cs\'a{}nyi}(2021)}]{Behler2021}%
  \BibitemOpen
  \bibfield  {author} {\bibinfo {author} {\bibfnamefont {J.}~\bibnamefont
  {Behler}}\ and\ \bibinfo {author} {\bibfnamefont {G.}~\bibnamefont
  {Cs\'a{}nyi}},\ }\bibfield  {title} {\enquote {\bibinfo {title} {Machine
  learning potentials for extended systems: {A} perspective},}\ }\href
  {\doibase 10.1140/epjb/s10051-021-00156-1} {\bibfield  {journal} {\bibinfo
  {journal} {Eur. Phys. J. B}\ }\textbf {\bibinfo {volume} {94}},\ \bibinfo
  {pages} {142} (\bibinfo {year} {2021})}\BibitemShut {NoStop}%
\bibitem [{\citenamefont {Morrow}\ and\ \citenamefont
  {Deringer}(2022)}]{Morrow2022}%
  \BibitemOpen
  \bibfield  {author} {\bibinfo {author} {\bibfnamefont {J.~D.}\ \bibnamefont
  {Morrow}}\ and\ \bibinfo {author} {\bibfnamefont {V.~L.}\ \bibnamefont
  {Deringer}},\ }\bibfield  {title} {\enquote {\bibinfo {title} {Indirect
  learning and physically guided validation of interatomic potential models},}\
  }\href@noop {} {\bibfield  {journal} {\bibinfo  {journal} {J. Chem. Phys.}\
  }\textbf {\bibinfo {volume} {157}},\ \bibinfo {pages} {104105} (\bibinfo
  {year} {2022})}\BibitemShut {NoStop}%
\bibitem [{\citenamefont {Pozdnyakov}\ \emph {et~al.}(2020)\citenamefont
  {Pozdnyakov}, \citenamefont {Willatt}, \citenamefont {Bart\'ok},
  \citenamefont {Ortner}, \citenamefont {Cs\'anyi},\ and\ \citenamefont
  {Ceriotti}}]{Pozdnyakov2020}%
  \BibitemOpen
  \bibfield  {author} {\bibinfo {author} {\bibfnamefont {S.~N.}\ \bibnamefont
  {Pozdnyakov}}, \bibinfo {author} {\bibfnamefont {M.~J.}\ \bibnamefont
  {Willatt}}, \bibinfo {author} {\bibfnamefont {A.~P.}\ \bibnamefont
  {Bart\'ok}}, \bibinfo {author} {\bibfnamefont {C.}~\bibnamefont {Ortner}},
  \bibinfo {author} {\bibfnamefont {G.}~\bibnamefont {Cs\'anyi}}, \ and\
  \bibinfo {author} {\bibfnamefont {M.}~\bibnamefont {Ceriotti}},\ }\bibfield
  {title} {\enquote {\bibinfo {title} {Incompleteness of atomic structure
  representations},}\ }\href {\doibase 10.1103/PhysRevLett.125.166001}
  {\bibfield  {journal} {\bibinfo  {journal} {Phys. Rev. Lett.}\ }\textbf
  {\bibinfo {volume} {125}},\ \bibinfo {pages} {166001} (\bibinfo {year}
  {2020})}\BibitemShut {NoStop}%
\bibitem [{\citenamefont {Behler}(2011)}]{Behler2011}%
  \BibitemOpen
  \bibfield  {author} {\bibinfo {author} {\bibfnamefont {J.}~\bibnamefont
  {Behler}},\ }\bibfield  {title} {\enquote {\bibinfo {title} {Atom-centered
  symmetry functions for constructing high-dimensional neural network
  potentials},}\ }\href {\doibase 10.1063/1.3553717} {\bibfield  {journal}
  {\bibinfo  {journal} {J. Chem. Phys.}\ }\textbf {\bibinfo {volume} {134}},\
  \bibinfo {pages} {074106} (\bibinfo {year} {2011})}\BibitemShut {NoStop}%
\bibitem [{\citenamefont {Bart\'o{}k}, \citenamefont {Kondor},\ and\
  \citenamefont {Cs\'a{}nyi}(2013)}]{Bartok2013}%
  \BibitemOpen
  \bibfield  {author} {\bibinfo {author} {\bibfnamefont {A.~P.}\ \bibnamefont
  {Bart\'o{}k}}, \bibinfo {author} {\bibfnamefont {R.}~\bibnamefont {Kondor}},
  \ and\ \bibinfo {author} {\bibfnamefont {G.}~\bibnamefont {Cs\'a{}nyi}},\
  }\bibfield  {title} {\enquote {\bibinfo {title} {On representing chemical
  environments},}\ }\href {\doibase 10.1103/PhysRevB.87.184115} {\bibfield
  {journal} {\bibinfo  {journal} {Phys. Rev. B}\ }\textbf {\bibinfo {volume}
  {87}},\ \bibinfo {pages} {184115} (\bibinfo {year} {2013})}\BibitemShut
  {NoStop}%
\bibitem [{\citenamefont {Thompson}\ \emph {et~al.}(2015)\citenamefont
  {Thompson}, \citenamefont {Swiler}, \citenamefont {Trott}, \citenamefont
  {Foiles},\ and\ \citenamefont {Tucker}}]{Thompson2015}%
  \BibitemOpen
  \bibfield  {author} {\bibinfo {author} {\bibfnamefont {A.~P.}\ \bibnamefont
  {Thompson}}, \bibinfo {author} {\bibfnamefont {L.~P.}\ \bibnamefont
  {Swiler}}, \bibinfo {author} {\bibfnamefont {C.~R.}\ \bibnamefont {Trott}},
  \bibinfo {author} {\bibfnamefont {S.~M.}\ \bibnamefont {Foiles}}, \ and\
  \bibinfo {author} {\bibfnamefont {G.~J.}\ \bibnamefont {Tucker}},\ }\bibfield
   {title} {\enquote {\bibinfo {title} {Spectral neighbor analysis method for
  automated generation of quantum-accurate interatomic potentials},}\ }\href
  {\doibase 10.1016/j.jcp.2014.12.018} {\bibfield  {journal} {\bibinfo
  {journal} {J. Comput. Phys.}\ }\textbf {\bibinfo {volume} {285}},\ \bibinfo
  {pages} {316--330} (\bibinfo {year} {2015})}\BibitemShut {NoStop}%
\bibitem [{\citenamefont {Sch\"u{}tt}\ \emph {et~al.}(2018)\citenamefont
  {Sch\"u{}tt}, \citenamefont {Sauceda}, \citenamefont {Kindermans},
  \citenamefont {Tkatchenko},\ and\ \citenamefont {M\"u{}ller}}]{Schuett2018}%
  \BibitemOpen
  \bibfield  {author} {\bibinfo {author} {\bibfnamefont {K.~T.}\ \bibnamefont
  {Sch\"u{}tt}}, \bibinfo {author} {\bibfnamefont {H.~E.}\ \bibnamefont
  {Sauceda}}, \bibinfo {author} {\bibfnamefont {P.-J.}\ \bibnamefont
  {Kindermans}}, \bibinfo {author} {\bibfnamefont {A.}~\bibnamefont
  {Tkatchenko}}, \ and\ \bibinfo {author} {\bibfnamefont {K.-R.}\ \bibnamefont
  {M\"u{}ller}},\ }\bibfield  {title} {\enquote {\bibinfo {title} {{SchNet} --
  a deep learning architecture for molecules and materials},}\ }\href {\doibase
  10.1063/1.5019779} {\bibfield  {journal} {\bibinfo  {journal} {J. Chem.
  Phys.}\ }\textbf {\bibinfo {volume} {148}},\ \bibinfo {pages} {241722}
  (\bibinfo {year} {2018})}\BibitemShut {NoStop}%
\bibitem [{\citenamefont {Batzner}\ \emph {et~al.}(2022)\citenamefont
  {Batzner}, \citenamefont {Musaelian}, \citenamefont {Sun}, \citenamefont
  {Geiger}, \citenamefont {Mailoa}, \citenamefont {Kornbluth}, \citenamefont
  {Molinari}, \citenamefont {Smidt},\ and\ \citenamefont
  {Kozinsky}}]{Batzner2022}%
  \BibitemOpen
  \bibfield  {author} {\bibinfo {author} {\bibfnamefont {S.}~\bibnamefont
  {Batzner}}, \bibinfo {author} {\bibfnamefont {A.}~\bibnamefont {Musaelian}},
  \bibinfo {author} {\bibfnamefont {L.}~\bibnamefont {Sun}}, \bibinfo {author}
  {\bibfnamefont {M.}~\bibnamefont {Geiger}}, \bibinfo {author} {\bibfnamefont
  {J.~P.}\ \bibnamefont {Mailoa}}, \bibinfo {author} {\bibfnamefont
  {M.}~\bibnamefont {Kornbluth}}, \bibinfo {author} {\bibfnamefont
  {N.}~\bibnamefont {Molinari}}, \bibinfo {author} {\bibfnamefont {T.~E.}\
  \bibnamefont {Smidt}}, \ and\ \bibinfo {author} {\bibfnamefont
  {B.}~\bibnamefont {Kozinsky}},\ }\bibfield  {title} {\enquote {\bibinfo
  {title} {E(3)-equivariant graph neural networks for data-efficient and
  accurate interatomic potentials},}\ }\href {\doibase
  10.1038/s41467-022-29939-5} {\bibfield  {journal} {\bibinfo  {journal} {Nat.
  Commun.}\ }\textbf {\bibinfo {volume} {13}},\ \bibinfo {pages} {2453}
  (\bibinfo {year} {2022})}\BibitemShut {NoStop}%
\bibitem [{\citenamefont {Musil}\ \emph {et~al.}(2021)\citenamefont {Musil},
  \citenamefont {Grisafi}, \citenamefont {Bart\'o{}k}, \citenamefont {Ortner},
  \citenamefont {Cs\'a{}nyi},\ and\ \citenamefont {Ceriotti}}]{Musil2021}%
  \BibitemOpen
  \bibfield  {author} {\bibinfo {author} {\bibfnamefont {F.}~\bibnamefont
  {Musil}}, \bibinfo {author} {\bibfnamefont {A.}~\bibnamefont {Grisafi}},
  \bibinfo {author} {\bibfnamefont {A.~P.}\ \bibnamefont {Bart\'o{}k}},
  \bibinfo {author} {\bibfnamefont {C.}~\bibnamefont {Ortner}}, \bibinfo
  {author} {\bibfnamefont {G.}~\bibnamefont {Cs\'a{}nyi}}, \ and\ \bibinfo
  {author} {\bibfnamefont {M.}~\bibnamefont {Ceriotti}},\ }\bibfield  {title}
  {\enquote {\bibinfo {title} {Physics-inspired structural representations for
  molecules and materials},}\ }\href {\doibase 10.1021/acs.chemrev.1c00021}
  {\bibfield  {journal} {\bibinfo  {journal} {Chem. Rev.}\ }\textbf {\bibinfo
  {volume} {121}},\ \bibinfo {pages} {9759--9815} (\bibinfo {year}
  {2021})}\BibitemShut {NoStop}%
\bibitem [{\citenamefont {Glielmo}, \citenamefont {Zeni},\ and\ \citenamefont
  {De~Vita}(2018)}]{Glielmo2018}%
  \BibitemOpen
  \bibfield  {author} {\bibinfo {author} {\bibfnamefont {A.}~\bibnamefont
  {Glielmo}}, \bibinfo {author} {\bibfnamefont {C.}~\bibnamefont {Zeni}}, \
  and\ \bibinfo {author} {\bibfnamefont {A.}~\bibnamefont {De~Vita}},\
  }\bibfield  {title} {\enquote {\bibinfo {title} {Efficient nonparametric
  $n$-body force fields from machine learning},}\ }\href {\doibase
  10.1103/PhysRevB.97.184307} {\bibfield  {journal} {\bibinfo  {journal} {Phys.
  Rev. B}\ }\textbf {\bibinfo {volume} {97}},\ \bibinfo {pages} {184307}
  (\bibinfo {year} {2018})}\BibitemShut {NoStop}%
\bibitem [{\citenamefont {Deringer}, \citenamefont {Pickard},\ and\
  \citenamefont {Cs\'anyi}(2018)}]{Deringer2018}%
  \BibitemOpen
  \bibfield  {author} {\bibinfo {author} {\bibfnamefont {V.~L.}\ \bibnamefont
  {Deringer}}, \bibinfo {author} {\bibfnamefont {C.~J.}\ \bibnamefont
  {Pickard}}, \ and\ \bibinfo {author} {\bibfnamefont {G.}~\bibnamefont
  {Cs\'anyi}},\ }\bibfield  {title} {\enquote {\bibinfo {title} {Data-driven
  learning of total and local energies in elemental boron},}\ }\href {\doibase
  10.1103/PhysRevLett.120.156001} {\bibfield  {journal} {\bibinfo  {journal}
  {Phys. Rev. Lett.}\ }\textbf {\bibinfo {volume} {120}},\ \bibinfo {pages}
  {156001} (\bibinfo {year} {2018})}\BibitemShut {NoStop}%
\bibitem [{\citenamefont {Behler}\ and\ \citenamefont
  {Parrinello}(2007)}]{Behler2007}%
  \BibitemOpen
  \bibfield  {author} {\bibinfo {author} {\bibfnamefont {J.}~\bibnamefont
  {Behler}}\ and\ \bibinfo {author} {\bibfnamefont {M.}~\bibnamefont
  {Parrinello}},\ }\bibfield  {title} {\enquote {\bibinfo {title} {Generalized
  neural-network representation of high-dimensional potential-energy
  surfaces},}\ }\href {\doibase 10.1103/PhysRevLett.98.146401} {\bibfield
  {journal} {\bibinfo  {journal} {Phys. Rev. Lett.}\ }\textbf {\bibinfo
  {volume} {98}},\ \bibinfo {pages} {146401} (\bibinfo {year}
  {2007})}\BibitemShut {NoStop}%
\bibitem [{\citenamefont {Artrith}\ and\ \citenamefont
  {Urban}(2016)}]{Artrith2016}%
  \BibitemOpen
  \bibfield  {author} {\bibinfo {author} {\bibfnamefont {N.}~\bibnamefont
  {Artrith}}\ and\ \bibinfo {author} {\bibfnamefont {A.}~\bibnamefont
  {Urban}},\ }\bibfield  {title} {\enquote {\bibinfo {title} {An implementation
  of artificial neural-network potentials for atomistic materials simulations:
  {Performance} for {TiO}$_{2}$},}\ }\href {\doibase
  10.1016/j.commatsci.2015.11.047} {\bibfield  {journal} {\bibinfo  {journal}
  {Comput. Mater. Sci.}\ }\textbf {\bibinfo {volume} {114}},\ \bibinfo {pages}
  {135--150} (\bibinfo {year} {2016})}\BibitemShut {NoStop}%
\bibitem [{\citenamefont {Smith}, \citenamefont {Isayev},\ and\ \citenamefont
  {Roitberg}(2017)}]{Smith2017}%
  \BibitemOpen
  \bibfield  {author} {\bibinfo {author} {\bibfnamefont {J.~S.}\ \bibnamefont
  {Smith}}, \bibinfo {author} {\bibfnamefont {O.}~\bibnamefont {Isayev}}, \
  and\ \bibinfo {author} {\bibfnamefont {A.~E.}\ \bibnamefont {Roitberg}},\
  }\bibfield  {title} {\enquote {\bibinfo {title} {Ani-1: {An} extensible
  neural network potential with {DFT} accuracy at force field computational
  cost},}\ }\href {\doibase 10.1039/C6SC05720A} {\bibfield  {journal} {\bibinfo
   {journal} {Chem. Sci.}\ }\textbf {\bibinfo {volume} {8}},\ \bibinfo {pages}
  {3192--3203} (\bibinfo {year} {2017})}\BibitemShut {NoStop}%
\bibitem [{\citenamefont {Zhang}\ \emph {et~al.}(2018)\citenamefont {Zhang},
  \citenamefont {Han}, \citenamefont {Wang}, \citenamefont {Car},\ and\
  \citenamefont {E}}]{Zhang2018a}%
  \BibitemOpen
  \bibfield  {author} {\bibinfo {author} {\bibfnamefont {L.}~\bibnamefont
  {Zhang}}, \bibinfo {author} {\bibfnamefont {J.}~\bibnamefont {Han}}, \bibinfo
  {author} {\bibfnamefont {H.}~\bibnamefont {Wang}}, \bibinfo {author}
  {\bibfnamefont {R.}~\bibnamefont {Car}}, \ and\ \bibinfo {author}
  {\bibfnamefont {W.}~\bibnamefont {E}},\ }\bibfield  {title} {\enquote
  {\bibinfo {title} {Deep potential molecular dynamics: {A} scalable model with
  the accuracy of quantum mechanics},}\ }\href {\doibase
  10.1103/PhysRevLett.120.143001} {\bibfield  {journal} {\bibinfo  {journal}
  {Phys. Rev. Lett.}\ }\textbf {\bibinfo {volume} {120}},\ \bibinfo {pages}
  {143001} (\bibinfo {year} {2018})}\BibitemShut {NoStop}%
\bibitem [{\citenamefont {Bart\'ok}\ \emph {et~al.}(2010)\citenamefont
  {Bart\'ok}, \citenamefont {Payne}, \citenamefont {Kondor},\ and\
  \citenamefont {Cs\'anyi}}]{Bartok2010}%
  \BibitemOpen
  \bibfield  {author} {\bibinfo {author} {\bibfnamefont {A.~P.}\ \bibnamefont
  {Bart\'ok}}, \bibinfo {author} {\bibfnamefont {M.~C.}\ \bibnamefont {Payne}},
  \bibinfo {author} {\bibfnamefont {R.}~\bibnamefont {Kondor}}, \ and\ \bibinfo
  {author} {\bibfnamefont {G.}~\bibnamefont {Cs\'anyi}},\ }\bibfield  {title}
  {\enquote {\bibinfo {title} {Gaussian approximation potentials: The accuracy
  of quantum mechanics, without the electrons},}\ }\href {\doibase
  10.1103/PhysRevLett.104.136403} {\bibfield  {journal} {\bibinfo  {journal}
  {Phys. Rev. Lett.}\ }\textbf {\bibinfo {volume} {104}},\ \bibinfo {pages}
  {136403} (\bibinfo {year} {2010})}\BibitemShut {NoStop}%
\bibitem [{\citenamefont {Shapeev}(2016)}]{Shapeev2016}%
  \BibitemOpen
  \bibfield  {author} {\bibinfo {author} {\bibfnamefont {A.}~\bibnamefont
  {Shapeev}},\ }\bibfield  {title} {\enquote {\bibinfo {title} {Moment tensor
  potentials: {A} class of systematically improvable interatomic potentials},}\
  }\href {\doibase 10.1137/15M1054183} {\bibfield  {journal} {\bibinfo
  {journal} {Multiscale Model. Simul.}\ }\textbf {\bibinfo {volume} {14}},\
  \bibinfo {pages} {1153--1173} (\bibinfo {year} {2016})}\BibitemShut {NoStop}%
\bibitem [{\citenamefont {Drautz}(2019)}]{Drautz2019}%
  \BibitemOpen
  \bibfield  {author} {\bibinfo {author} {\bibfnamefont {R.}~\bibnamefont
  {Drautz}},\ }\bibfield  {title} {\enquote {\bibinfo {title} {Atomic cluster
  expansion for accurate and transferable interatomic potentials},}\ }\href
  {\doibase 10.1103/PhysRevB.99.014104} {\bibfield  {journal} {\bibinfo
  {journal} {Phys. Rev. B}\ }\textbf {\bibinfo {volume} {99}},\ \bibinfo
  {pages} {014104} (\bibinfo {year} {2019})}\BibitemShut {NoStop}%
\bibitem [{\citenamefont {Bochkarev}\ \emph {et~al.}(2022)\citenamefont
  {Bochkarev}, \citenamefont {Lysogorskiy}, \citenamefont {Ortner},
  \citenamefont {Cs\'anyi},\ and\ \citenamefont {Drautz}}]{Bochkarev2022}%
  \BibitemOpen
  \bibfield  {author} {\bibinfo {author} {\bibfnamefont {A.}~\bibnamefont
  {Bochkarev}}, \bibinfo {author} {\bibfnamefont {Y.}~\bibnamefont
  {Lysogorskiy}}, \bibinfo {author} {\bibfnamefont {C.}~\bibnamefont {Ortner}},
  \bibinfo {author} {\bibfnamefont {G.}~\bibnamefont {Cs\'anyi}}, \ and\
  \bibinfo {author} {\bibfnamefont {R.}~\bibnamefont {Drautz}},\ }\bibfield
  {title} {\enquote {\bibinfo {title} {Multilayer atomic cluster expansion for
  semi-local interactions},}\ }\href {\doibase
  https://doi.org/10.48550/arXiv.2205.08177} {\bibfield  {journal} {\bibinfo
  {journal} {arXiv preprint\!\!}\ ,\ \bibinfo {pages} {arXiv:2205.08177
  [cond--mat.mtrl--sci]}} (\bibinfo {year} {2022})}\BibitemShut {NoStop}%
\bibitem [{\citenamefont {Behler}(2021)}]{Behler2021a}%
  \BibitemOpen
  \bibfield  {author} {\bibinfo {author} {\bibfnamefont {J.}~\bibnamefont
  {Behler}},\ }\bibfield  {title} {\enquote {\bibinfo {title} {Four generations
  of high-dimensional neural network potentials},}\ }\href {\doibase
  10.1021/acs.chemrev.0c00868} {\bibfield  {journal} {\bibinfo  {journal}
  {Chem. Rev.}\ }\textbf {\bibinfo {volume} {121}},\ \bibinfo {pages}
  {10037--10072} (\bibinfo {year} {2021})}\BibitemShut {NoStop}%
\bibitem [{\citenamefont {Pinheiro}\ \emph {et~al.}(2021)\citenamefont
  {Pinheiro}, \citenamefont {Ge}, \citenamefont {Ferr\'e{}}, \citenamefont
  {Dral},\ and\ \citenamefont {Barbatti}}]{Pinheiro2021}%
  \BibitemOpen
  \bibfield  {author} {\bibinfo {author} {\bibfnamefont {M.}~\bibnamefont
  {Pinheiro}}, \bibinfo {author} {\bibfnamefont {F.}~\bibnamefont {Ge}},
  \bibinfo {author} {\bibfnamefont {N.}~\bibnamefont {Ferr\'e{}}}, \bibinfo
  {author} {\bibfnamefont {P.~O.}\ \bibnamefont {Dral}}, \ and\ \bibinfo
  {author} {\bibfnamefont {M.}~\bibnamefont {Barbatti}},\ }\bibfield  {title}
  {\enquote {\bibinfo {title} {Choosing the right molecular machine learning
  potential},}\ }\href {\doibase 10.1039/D1SC03564A} {\bibfield  {journal}
  {\bibinfo  {journal} {Chem. Sci.}\ }\textbf {\bibinfo {volume} {12}},\
  \bibinfo {pages} {14396--14413} (\bibinfo {year} {2021})}\BibitemShut
  {NoStop}%
\bibitem [{\citenamefont {Deringer}\ and\ \citenamefont
  {Cs\'anyi}(2017)}]{Deringer2017}%
  \BibitemOpen
  \bibfield  {author} {\bibinfo {author} {\bibfnamefont {V.~L.}\ \bibnamefont
  {Deringer}}\ and\ \bibinfo {author} {\bibfnamefont {G.}~\bibnamefont
  {Cs\'anyi}},\ }\bibfield  {title} {\enquote {\bibinfo {title} {Machine
  learning based interatomic potential for amorphous carbon},}\ }\href
  {\doibase 10.1103/PhysRevB.95.094203} {\bibfield  {journal} {\bibinfo
  {journal} {Phys. Rev. B}\ }\textbf {\bibinfo {volume} {95}},\ \bibinfo
  {pages} {094203} (\bibinfo {year} {2017})}\BibitemShut {NoStop}%
\bibitem [{\citenamefont {Pickard}\ and\ \citenamefont
  {Needs}(2006)}]{Pickard2006}%
  \BibitemOpen
  \bibfield  {author} {\bibinfo {author} {\bibfnamefont {C.~J.}\ \bibnamefont
  {Pickard}}\ and\ \bibinfo {author} {\bibfnamefont {R.~J.}\ \bibnamefont
  {Needs}},\ }\bibfield  {title} {\enquote {\bibinfo {title} {High-pressure
  phases of silane},}\ }\href {\doibase 10.1103/PhysRevLett.97.045504}
  {\bibfield  {journal} {\bibinfo  {journal} {Phys. Rev. Lett.}\ }\textbf
  {\bibinfo {volume} {97}},\ \bibinfo {pages} {045504} (\bibinfo {year}
  {2006})}\BibitemShut {NoStop}%
\bibitem [{\citenamefont {Pickard}\ and\ \citenamefont
  {Needs}(2011)}]{Pickard2011}%
  \BibitemOpen
  \bibfield  {author} {\bibinfo {author} {\bibfnamefont {C.~J.}\ \bibnamefont
  {Pickard}}\ and\ \bibinfo {author} {\bibfnamefont {R.~J.}\ \bibnamefont
  {Needs}},\ }\bibfield  {title} {\enquote {\bibinfo {title} {Ab initio random
  structure searching},}\ }\href {\doibase 10.1088/0953-8984/23/5/053201}
  {\bibfield  {journal} {\bibinfo  {journal} {J. Phys.: Condens. Matter}\
  }\textbf {\bibinfo {volume} {23}},\ \bibinfo {pages} {053201} (\bibinfo
  {year} {2011})}\BibitemShut {NoStop}%
\bibitem [{\citenamefont {Bernstein}, \citenamefont {Cs\'a{}nyi},\ and\
  \citenamefont {Deringer}(2019)}]{Bernstein2019b}%
  \BibitemOpen
  \bibfield  {author} {\bibinfo {author} {\bibfnamefont {N.}~\bibnamefont
  {Bernstein}}, \bibinfo {author} {\bibfnamefont {G.}~\bibnamefont
  {Cs\'a{}nyi}}, \ and\ \bibinfo {author} {\bibfnamefont {V.~L.}\ \bibnamefont
  {Deringer}},\ }\bibfield  {title} {\enquote {\bibinfo {title} {De novo
  exploration and self-guided learning of potential-energy surfaces},}\ }\href
  {\doibase 10.1038/s41524-019-0236-6} {\bibfield  {journal} {\bibinfo
  {journal} {npj Comput. Mater.}\ }\textbf {\bibinfo {volume} {5}},\ \bibinfo
  {pages} {99} (\bibinfo {year} {2019})}\BibitemShut {NoStop}%
\bibitem [{\citenamefont {Bart\'ok}\ \emph {et~al.}(2018)\citenamefont
  {Bart\'ok}, \citenamefont {Kermode}, \citenamefont {Bernstein},\ and\
  \citenamefont {Cs\'anyi}}]{Bartok2018}%
  \BibitemOpen
  \bibfield  {author} {\bibinfo {author} {\bibfnamefont {A.~P.}\ \bibnamefont
  {Bart\'ok}}, \bibinfo {author} {\bibfnamefont {J.}~\bibnamefont {Kermode}},
  \bibinfo {author} {\bibfnamefont {N.}~\bibnamefont {Bernstein}}, \ and\
  \bibinfo {author} {\bibfnamefont {G.}~\bibnamefont {Cs\'anyi}},\ }\bibfield
  {title} {\enquote {\bibinfo {title} {Machine learning a general-purpose
  interatomic potential for silicon},}\ }\href {\doibase
  10.1103/PhysRevX.8.041048} {\bibfield  {journal} {\bibinfo  {journal} {Phys.
  Rev. X}\ }\textbf {\bibinfo {volume} {8}},\ \bibinfo {pages} {041048}
  (\bibinfo {year} {2018})}\BibitemShut {NoStop}%
\bibitem [{\citenamefont {Deringer}\ \emph {et~al.}(2018)\citenamefont
  {Deringer}, \citenamefont {Bernstein}, \citenamefont {Bart\'o{}k},
  \citenamefont {Cliffe}, \citenamefont {Kerber}, \citenamefont {Marbella},
  \citenamefont {Grey}, \citenamefont {Elliott},\ and\ \citenamefont
  {Cs\'a{}nyi}}]{Deringer2018d}%
  \BibitemOpen
  \bibfield  {author} {\bibinfo {author} {\bibfnamefont {V.~L.}\ \bibnamefont
  {Deringer}}, \bibinfo {author} {\bibfnamefont {N.}~\bibnamefont {Bernstein}},
  \bibinfo {author} {\bibfnamefont {A.~P.}\ \bibnamefont {Bart\'o{}k}},
  \bibinfo {author} {\bibfnamefont {M.~J.}\ \bibnamefont {Cliffe}}, \bibinfo
  {author} {\bibfnamefont {R.~N.}\ \bibnamefont {Kerber}}, \bibinfo {author}
  {\bibfnamefont {L.~E.}\ \bibnamefont {Marbella}}, \bibinfo {author}
  {\bibfnamefont {C.~P.}\ \bibnamefont {Grey}}, \bibinfo {author}
  {\bibfnamefont {S.~R.}\ \bibnamefont {Elliott}}, \ and\ \bibinfo {author}
  {\bibfnamefont {G.}~\bibnamefont {Cs\'a{}nyi}},\ }\bibfield  {title}
  {\enquote {\bibinfo {title} {Realistic atomistic structure of amorphous
  silicon from machine-learning-driven molecular dynamics},}\ }\href {\doibase
  10.1021/acs.jpclett.8b00902} {\bibfield  {journal} {\bibinfo  {journal} {J.
  Phys. Chem. Lett.}\ }\textbf {\bibinfo {volume} {9}},\ \bibinfo {pages}
  {2879--2885} (\bibinfo {year} {2018})}\BibitemShut {NoStop}%
\bibitem [{\citenamefont {Stillinger}\ and\ \citenamefont
  {Weber}(1985)}]{Stillinger1985}%
  \BibitemOpen
  \bibfield  {author} {\bibinfo {author} {\bibfnamefont {F.~H.}\ \bibnamefont
  {Stillinger}}\ and\ \bibinfo {author} {\bibfnamefont {T.~A.}\ \bibnamefont
  {Weber}},\ }\bibfield  {title} {\enquote {\bibinfo {title} {Computer
  simulation of local order in condensed phases of silicon},}\ }\href {\doibase
  10.1103/PhysRevB.31.5262} {\bibfield  {journal} {\bibinfo  {journal} {Phys.
  Rev. B}\ }\textbf {\bibinfo {volume} {31}},\ \bibinfo {pages} {5262--5271}
  (\bibinfo {year} {1985})}\BibitemShut {NoStop}%
\bibitem [{\citenamefont {Jain}\ \emph {et~al.}(2013)\citenamefont {Jain},
  \citenamefont {Ong}, \citenamefont {Hautier}, \citenamefont {Chen},
  \citenamefont {Richards}, \citenamefont {Dacek}, \citenamefont {Cholia},
  \citenamefont {Gunter}, \citenamefont {Skinner}, \citenamefont {Ceder},\ and\
  \citenamefont {Persson}}]{Jain2013}%
  \BibitemOpen
  \bibfield  {author} {\bibinfo {author} {\bibfnamefont {A.}~\bibnamefont
  {Jain}}, \bibinfo {author} {\bibfnamefont {S.~P.}\ \bibnamefont {Ong}},
  \bibinfo {author} {\bibfnamefont {G.}~\bibnamefont {Hautier}}, \bibinfo
  {author} {\bibfnamefont {W.}~\bibnamefont {Chen}}, \bibinfo {author}
  {\bibfnamefont {W.~D.}\ \bibnamefont {Richards}}, \bibinfo {author}
  {\bibfnamefont {S.}~\bibnamefont {Dacek}}, \bibinfo {author} {\bibfnamefont
  {S.}~\bibnamefont {Cholia}}, \bibinfo {author} {\bibfnamefont
  {D.}~\bibnamefont {Gunter}}, \bibinfo {author} {\bibfnamefont
  {D.}~\bibnamefont {Skinner}}, \bibinfo {author} {\bibfnamefont
  {G.}~\bibnamefont {Ceder}}, \ and\ \bibinfo {author} {\bibfnamefont {K.~A.}\
  \bibnamefont {Persson}},\ }\bibfield  {title} {\enquote {\bibinfo {title}
  {Commentary: {The Materials Project}: {A} materials genome approach to
  accelerating materials innovation},}\ }\href {\doibase 10.1063/1.4812323}
  {\bibfield  {journal} {\bibinfo  {journal} {APL Mater.}\ }\textbf {\bibinfo
  {volume} {1}},\ \bibinfo {pages} {011002} (\bibinfo {year}
  {2013})}\BibitemShut {NoStop}%
\bibitem [{\citenamefont {Pickard}(2022)}]{Pickard2022}%
  \BibitemOpen
  \bibfield  {author} {\bibinfo {author} {\bibfnamefont {C.~J.}\ \bibnamefont
  {Pickard}},\ }\bibfield  {title} {\enquote {\bibinfo {title} {Ephemeral data
  derived potentials for random structure search},}\ }\href {\doibase
  10.1103/PhysRevB.106.014102} {\bibfield  {journal} {\bibinfo  {journal}
  {Phys. Rev. B}\ }\textbf {\bibinfo {volume} {106}},\ \bibinfo {pages}
  {014102} (\bibinfo {year} {2022})}\BibitemShut {NoStop}%
\bibitem [{\citenamefont {Zagorac}\ \emph {et~al.}(2019)\citenamefont
  {Zagorac}, \citenamefont {M\"u{}ller}, \citenamefont {Ruehl}, \citenamefont
  {Zagorac},\ and\ \citenamefont {Rehme}}]{Zagorac2019}%
  \BibitemOpen
  \bibfield  {author} {\bibinfo {author} {\bibfnamefont {D.}~\bibnamefont
  {Zagorac}}, \bibinfo {author} {\bibfnamefont {H.}~\bibnamefont {M\"u{}ller}},
  \bibinfo {author} {\bibfnamefont {S.}~\bibnamefont {Ruehl}}, \bibinfo
  {author} {\bibfnamefont {J.}~\bibnamefont {Zagorac}}, \ and\ \bibinfo
  {author} {\bibfnamefont {S.}~\bibnamefont {Rehme}},\ }\bibfield  {title}
  {\enquote {\bibinfo {title} {Recent developments in the {Inorganic} {Crystal}
  {Structure} {Database}: theoretical crystal structure data and related
  features},}\ }\href {\doibase 10.1107/S160057671900997X} {\bibfield
  {journal} {\bibinfo  {journal} {J. Appl. Crystallogr.}\ }\textbf {\bibinfo
  {volume} {52}},\ \bibinfo {pages} {918--925} (\bibinfo {year}
  {2019})}\BibitemShut {NoStop}%
\bibitem [{\citenamefont {Erhard}\ \emph {et~al.}(2022)\citenamefont {Erhard},
  \citenamefont {Rohrer}, \citenamefont {Albe},\ and\ \citenamefont
  {Deringer}}]{Erhard2022}%
  \BibitemOpen
  \bibfield  {author} {\bibinfo {author} {\bibfnamefont {L.~C.}\ \bibnamefont
  {Erhard}}, \bibinfo {author} {\bibfnamefont {J.}~\bibnamefont {Rohrer}},
  \bibinfo {author} {\bibfnamefont {K.}~\bibnamefont {Albe}}, \ and\ \bibinfo
  {author} {\bibfnamefont {V.~L.}\ \bibnamefont {Deringer}},\ }\bibfield
  {title} {\enquote {\bibinfo {title} {A machine-learned interatomic potential
  for silica and its relation to empirical models},}\ }\href {\doibase
  10.1038/s41524-022-00768-w} {\bibfield  {journal} {\bibinfo  {journal} {npj
  Comput. Mater.}\ }\textbf {\bibinfo {volume} {8}},\ \bibinfo {pages} {90}
  (\bibinfo {year} {2022})}\BibitemShut {NoStop}%
\bibitem [{\citenamefont {Togo}\ and\ \citenamefont
  {Tanaka}(2015)}]{Togo2015a}%
  \BibitemOpen
  \bibfield  {author} {\bibinfo {author} {\bibfnamefont {A.}~\bibnamefont
  {Togo}}\ and\ \bibinfo {author} {\bibfnamefont {I.}~\bibnamefont {Tanaka}},\
  }\bibfield  {title} {\enquote {\bibinfo {title} {First principles phonon
  calculations in materials science},}\ }\href {\doibase
  10.1016/j.scriptamat.2015.07.021} {\bibfield  {journal} {\bibinfo  {journal}
  {Scripta Mater.}\ }\textbf {\bibinfo {volume} {108}},\ \bibinfo {pages}
  {1--5} (\bibinfo {year} {2015})}\BibitemShut {NoStop}%
\bibitem [{\citenamefont {Gastegger}, \citenamefont {Behler},\ and\
  \citenamefont {Marquetand}(2017)}]{Gastegger2017}%
  \BibitemOpen
  \bibfield  {author} {\bibinfo {author} {\bibfnamefont {M.}~\bibnamefont
  {Gastegger}}, \bibinfo {author} {\bibfnamefont {J.}~\bibnamefont {Behler}}, \
  and\ \bibinfo {author} {\bibfnamefont {P.}~\bibnamefont {Marquetand}},\
  }\bibfield  {title} {\enquote {\bibinfo {title} {Machine learning molecular
  dynamics for the simulation of infrared spectra},}\ }\href {\doibase
  10.1039/C7SC02267K} {\bibfield  {journal} {\bibinfo  {journal} {Chem. Sci.}\
  }\textbf {\bibinfo {volume} {8}},\ \bibinfo {pages} {6924--6935} (\bibinfo
  {year} {2017})}\BibitemShut {NoStop}%
\bibitem [{\citenamefont {Westermayr}\ \emph {et~al.}(2021)\citenamefont
  {Westermayr}, \citenamefont {Gastegger}, \citenamefont {Sch\"u{}tt},\ and\
  \citenamefont {Maurer}}]{Westermayr2021}%
  \BibitemOpen
  \bibfield  {author} {\bibinfo {author} {\bibfnamefont {J.}~\bibnamefont
  {Westermayr}}, \bibinfo {author} {\bibfnamefont {M.}~\bibnamefont
  {Gastegger}}, \bibinfo {author} {\bibfnamefont {K.~T.}\ \bibnamefont
  {Sch\"u{}tt}}, \ and\ \bibinfo {author} {\bibfnamefont {R.~J.}\ \bibnamefont
  {Maurer}},\ }\bibfield  {title} {\enquote {\bibinfo {title} {Perspective on
  integrating machine learning into computational chemistry and materials
  science},}\ }\href {\doibase 10.1063/5.0047760} {\bibfield  {journal}
  {\bibinfo  {journal} {J. Chem. Phys.}\ }\textbf {\bibinfo {volume} {154}},\
  \bibinfo {pages} {230903} (\bibinfo {year} {2021})}\BibitemShut {NoStop}%
\bibitem [{\citenamefont {Ceriotti}(2022)}]{Ceriotti2022}%
  \BibitemOpen
  \bibfield  {author} {\bibinfo {author} {\bibfnamefont {M.}~\bibnamefont
  {Ceriotti}},\ }\bibfield  {title} {\enquote {\bibinfo {title} {Beyond
  potentials: integrated machine-learning models for materials},}\ }\href
  {\doibase https://doi.org/10.48550/arXiv.2208.06139} {\bibfield  {journal}
  {\bibinfo  {journal} {arXiv preprint\!\!}\ ,\ \bibinfo {pages}
  {arXiv:2208.06139 [cond--mat.mtrl--sci]}} (\bibinfo {year}
  {2022})}\BibitemShut {NoStop}%
\bibitem [{\citenamefont {Golze}\ \emph {et~al.}(2022)\citenamefont {Golze},
  \citenamefont {Hirvensalo}, \citenamefont {Hern\'a{}ndez-Le\'o{}n},
  \citenamefont {Aarva}, \citenamefont {Etula}, \citenamefont {Susi},
  \citenamefont {Rinke}, \citenamefont {Laurila},\ and\ \citenamefont
  {Caro}}]{Golze2022}%
  \BibitemOpen
  \bibfield  {author} {\bibinfo {author} {\bibfnamefont {D.}~\bibnamefont
  {Golze}}, \bibinfo {author} {\bibfnamefont {M.}~\bibnamefont {Hirvensalo}},
  \bibinfo {author} {\bibfnamefont {P.}~\bibnamefont {Hern\'a{}ndez-Le\'o{}n}},
  \bibinfo {author} {\bibfnamefont {A.}~\bibnamefont {Aarva}}, \bibinfo
  {author} {\bibfnamefont {J.}~\bibnamefont {Etula}}, \bibinfo {author}
  {\bibfnamefont {T.}~\bibnamefont {Susi}}, \bibinfo {author} {\bibfnamefont
  {P.}~\bibnamefont {Rinke}}, \bibinfo {author} {\bibfnamefont
  {T.}~\bibnamefont {Laurila}}, \ and\ \bibinfo {author} {\bibfnamefont
  {M.~A.}\ \bibnamefont {Caro}},\ }\bibfield  {title} {\enquote {\bibinfo
  {title} {Accurate computational prediction of core-electron binding energies
  in carbon-based materials: {A} machine-learning model combining
  density-functional theory and {GW}},}\ }\href {\doibase
  10.1021/acs.chemmater.1c04279} {\bibfield  {journal} {\bibinfo  {journal}
  {Chem. Mater.}\ }\textbf {\bibinfo {volume} {34}},\ \bibinfo {pages}
  {6240--6254} (\bibinfo {year} {2022})}\BibitemShut {NoStop}%
\bibitem [{\citenamefont {Shapeev}, \citenamefont {Bocharov},\ and\
  \citenamefont {Kuzmin}(2022)}]{Shapeev2022}%
  \BibitemOpen
  \bibfield  {author} {\bibinfo {author} {\bibfnamefont {A.~V.}\ \bibnamefont
  {Shapeev}}, \bibinfo {author} {\bibfnamefont {D.}~\bibnamefont {Bocharov}}, \
  and\ \bibinfo {author} {\bibfnamefont {A.}~\bibnamefont {Kuzmin}},\
  }\bibfield  {title} {\enquote {\bibinfo {title} {Validation of moment tensor
  potentials for fcc and bcc metals using {EXAFS} spectra},}\ }\href {\doibase
  10.1016/j.commatsci.2021.111028} {\bibfield  {journal} {\bibinfo  {journal}
  {Comput. Mater. Sci.}\ }\textbf {\bibinfo {volume} {210}},\ \bibinfo {pages}
  {111028} (\bibinfo {year} {2022})}\BibitemShut {NoStop}%
\bibitem [{\citenamefont {Forse}\ \emph {et~al.}(2015)\citenamefont {Forse},
  \citenamefont {Merlet}, \citenamefont {Allan}, \citenamefont {Humphreys},
  \citenamefont {Griffin}, \citenamefont {Aslan}, \citenamefont {Zeiger},
  \citenamefont {Presser}, \citenamefont {Gogotsi},\ and\ \citenamefont
  {Grey}}]{Forse2015}%
  \BibitemOpen
  \bibfield  {author} {\bibinfo {author} {\bibfnamefont {A.~C.}\ \bibnamefont
  {Forse}}, \bibinfo {author} {\bibfnamefont {C.}~\bibnamefont {Merlet}},
  \bibinfo {author} {\bibfnamefont {P.~K.}\ \bibnamefont {Allan}}, \bibinfo
  {author} {\bibfnamefont {E.~K.}\ \bibnamefont {Humphreys}}, \bibinfo {author}
  {\bibfnamefont {J.~M.}\ \bibnamefont {Griffin}}, \bibinfo {author}
  {\bibfnamefont {M.}~\bibnamefont {Aslan}}, \bibinfo {author} {\bibfnamefont
  {M.}~\bibnamefont {Zeiger}}, \bibinfo {author} {\bibfnamefont
  {V.}~\bibnamefont {Presser}}, \bibinfo {author} {\bibfnamefont
  {Y.}~\bibnamefont {Gogotsi}}, \ and\ \bibinfo {author} {\bibfnamefont
  {C.~P.}\ \bibnamefont {Grey}},\ }\bibfield  {title} {\enquote {\bibinfo
  {title} {New insights into the structure of nanoporous carbons from {NMR},
  {Raman}, and pair distribution function analysis},}\ }\href {\doibase
  10.1021/acs.chemmater.5b03216} {\bibfield  {journal} {\bibinfo  {journal}
  {Chem. Mater.}\ }\textbf {\bibinfo {volume} {27}},\ \bibinfo {pages}
  {6848--6857} (\bibinfo {year} {2015})}\BibitemShut {NoStop}%
\bibitem [{\citenamefont {Wang}\ \emph
  {et~al.}(2022{\natexlab{b}})\citenamefont {Wang}, \citenamefont {Fan},
  \citenamefont {Qian}, \citenamefont {Ala-Nissila},\ and\ \citenamefont
  {Caro}}]{Wang2022a}%
  \BibitemOpen
  \bibfield  {author} {\bibinfo {author} {\bibfnamefont {Y.}~\bibnamefont
  {Wang}}, \bibinfo {author} {\bibfnamefont {Z.}~\bibnamefont {Fan}}, \bibinfo
  {author} {\bibfnamefont {P.}~\bibnamefont {Qian}}, \bibinfo {author}
  {\bibfnamefont {T.}~\bibnamefont {Ala-Nissila}}, \ and\ \bibinfo {author}
  {\bibfnamefont {M.~A.}\ \bibnamefont {Caro}},\ }\bibfield  {title} {\enquote
  {\bibinfo {title} {Structure and pore size distribution in nanoporous
  carbon},}\ }\href {\doibase 10.1021/acs.chemmater.1c03279} {\bibfield
  {journal} {\bibinfo  {journal} {Chem. Mater.}\ }\textbf {\bibinfo {volume}
  {34}},\ \bibinfo {pages} {617--628} (\bibinfo {year}
  {2022}{\natexlab{b}})}\BibitemShut {NoStop}%
\bibitem [{\citenamefont {Mata}\ and\ \citenamefont {Suhm}(2017)}]{Mata2017}%
  \BibitemOpen
  \bibfield  {author} {\bibinfo {author} {\bibfnamefont {R.~A.}\ \bibnamefont
  {Mata}}\ and\ \bibinfo {author} {\bibfnamefont {M.~A.}\ \bibnamefont
  {Suhm}},\ }\bibfield  {title} {\enquote {\bibinfo {title} {Benchmarking
  quantum chemical methods: Are we heading in the right direction?}}\ }\href
  {\doibase 10.1002/anie.201611308} {\bibfield  {journal} {\bibinfo  {journal}
  {Angew. Chem. Int. Ed.}\ }\textbf {\bibinfo {volume} {56}},\ \bibinfo {pages}
  {11011--11018} (\bibinfo {year} {2017})}\BibitemShut {NoStop}%
\bibitem [{\citenamefont {Stukowski}(2009)}]{Stukowski2009}%
  \BibitemOpen
  \bibfield  {author} {\bibinfo {author} {\bibfnamefont {A.}~\bibnamefont
  {Stukowski}},\ }\bibfield  {title} {\enquote {\bibinfo {title} {Visualization
  and analysis of atomistic simulation data with {OVITO}{\textendash}the {Open}
  {Visualization} {Tool}},}\ }\href {\doibase 10.1088/0965-0393/18/1/015012}
  {\bibfield  {journal} {\bibinfo  {journal} {Model. Simul. Mater. Sci. Eng.}\
  }\textbf {\bibinfo {volume} {18}},\ \bibinfo {pages} {015012} (\bibinfo
  {year} {2009})}\BibitemShut {NoStop}%
\end{thebibliography}
\end{document}